\documentclass[sigconf]{acmart}

\ifdefined\evenshorterappendix \def\shortappendix{} \else \fi
\def\arXiv{}

\usepackage{epsfig,endnotes}
\usepackage[utf8]{inputenc}

\usepackage{amsmath}
\usepackage{amsfonts}
\usepackage{amssymb}
\usepackage{amsthm}

\usepackage{bbm}
\usepackage{mathtools}
\usepackage[nounderscore]{syntax}
\usepackage{algorithm}
\usepackage{algpseudocode}
\usepackage{graphicx}

\usepackage{subcaption}
\usepackage{stmaryrd}
\usepackage{url}

\usepackage{standalone}
\usepackage{cancel}
\usepackage{listings}
\usepackage{xspace}

\usepackage{tikz} \usetikzlibrary{
  shapes,calc,arrows,fit,positioning,decorations.pathmorphing,snakes,intersections,shapes.geometric,trees
}

\newcommand{\pxm}   [1]{{\color{red}   {PM: #1}}}

\newcommand{\cut}[1]{ }

\newcommand{\dataset}[0]{\mathcal{D}}
\newcommand{\datasize}[0]{\dataset}
\newcommand{\prog}[0]{p}
\newcommand{\progsize}[0]{\prog}

\newcommand{\zs}[0]{Z}

\newcommand{\assocmeasure}[0]{\ensuremath{d}}
\newcommand{\infmeasure}  [0]{\ensuremath{\iota}}

\newcommand{\utilmeasure} [0]{\ensuremath{v}}

\newcommand{\proxyusage}   [0]{Proxy Use\xspace}

\newcommand{\argproxyusage}[2]{$(#1,#2)$-\proxyusage}
\newcommand{\edproxyusage} [0]{\argproxyusage{\epsilon}{\delta}}

\newcommand{\proxyuse}   [0]{proxy use\xspace}
\newcommand{\proxyuses}  [0]{proxy uses\xspace}
\newcommand{\argproxyuse}[2]{$(#1,#2)$-\proxyuse}
\newcommand{\edproxyuse} [0]{\argproxyuse{\epsilon}{\delta}}
\newcommand{\Proxyuse}   [0]{Proxy use\xspace}

\newcommand{\ProxyUse}   [0]{Proxy Use\xspace}

\newtheorem{thm}   {Theorem}
\newtheorem{defn}  {Definition}
\newtheorem{property}{Property}

\makeatletter
\def\thm@space@setup{\thm@preskip=1pt
\thm@postskip=1pt}
\makeatother

\newcommand{\code}[1]{\lstinline!#1!}

\newcommand{\stacklabel}[1]{\stackrel{\smash{\scriptscriptstyle \mathrm{#1}}}}
\newcommand{\defeq}{\stacklabel{def}=}

\newcommand{\bigO}[1]{\mathcal{O}\paren{#1}}

\newcommand{\paren}[1]{\left( #1 \right)}
\newcommand{\sparen}[1]{\left[ #1 \right]}
\newcommand{\set}[1]{\left\{ #1 \right\}}

\newcommand{\setsize}[1]{\left|#1\right|}

\DeclareMathOperator*{\E}{\mathbb{E}}
\newcommand{\prob}[1]{\Pr\paren{#1}}

\newcommand{\expectsub}[2]{\E_{#1}\sparen{#2}}

\newcommand{\support}[1]{\text{support}\paren{#1}}

\newcommand{\features}{\mathbf{X}}

\newcommand{\samp}{\ensuremath{\stacklabel{\$}\gets}}

\newcommand{\cal}{\mathcal}
\newcommand{\oracle}{{\cal O}}
\newcommand{\model}{{\cal A}}
\newcommand{\population}{{\cal P}}

\newcommand{\denote}[1]{\llbracket #1 \rrbracket}

\newcommand{\subst}[3]{[#1/#2]#3}
\newcommand{\pair}[2]{\langle #1, #2\rangle}
\newcommand{\pairpop}[2]{\pair{#1}{#2}_{\population}}
\newcommand{\pairpopbrace}[2]{\pair{\{#1\}}{#2}_{\population}}

\DeclareMathOperator*{\argmax}{arg\,max}
\DeclareMathAlphabet{\mathcal}{OMS}{cmsy}{m}{n}  %

\definecolor{mygray}{rgb}{0.5,0.5,0.5}
\lstdefinestyle{customlisp}{
  belowcaptionskip=1\baselineskip,
  breaklines=true,
  language=Lisp,
  showstringspaces=false,
  numbers=left,
  xleftmargin=2em,
  framexleftmargin=1.5em,
  numbersep=5pt,
  numberstyle=\tiny\color{mygray},
  basicstyle=\small\ttfamily,
  keywordstyle=\color{blue},
  commentstyle=\itshape\color{purple!40!black},
  stringstyle=\color{orange},
  morekeywords={def-rule, ?},
  tabsize=2
}

\makeatletter
\renewcommand{\paragraph}{%
  \@startsection{paragraph}{4}%
  {\z@}{0.75ex \@plus 0.5ex \@minus .2ex}{-1em}%
  {\normalfont\normalsize\bfseries}%
}
\makeatother

\usepackage{enumitem}
\setlist{nosep}

\algtext*{EndFor}
\algtext*{EndIf}
\algtext*{EndProcedure}

\begin{document}

\copyrightyear{2017}
\acmYear{2017}
\setcopyright{acmcopyright}
\acmConference{CCS '17}{October 30-November 3, 2017}{Dallas, TX, USA}\acmPrice{15.00}\acmDOI{http://dx.doi.org/10.1145/3133956.3134097}
\acmISBN{978-1-4503-4946-8/17/10}

\fancyhead{}
\settopmatter{printacmref=false,  printfolios=false} %

\title{Use Privacy in Data-Driven Systems}
\subtitle{Theory and Experiments with
  Machine Learnt Programs}

\author{Anupam Datta}
\affiliation{\institution{Carnegie Mellon University}}
\author{Matt Fredrikson}
\affiliation{\institution{Carnegie Mellon University}}
\author{Gihyuk Ko}
\affiliation{\institution{Carnegie Mellon University}}
\author{Piotr Mardziel}
\affiliation{\institution{Carnegie Mellon University}}
\author{Shayak Sen}
\affiliation{\institution{Carnegie Mellon University}}

\begin{abstract}
  This paper presents an approach to formalizing and enforcing
  a class of \emph{use privacy} properties in data-driven systems.
  In contrast to prior work, we focus on use restrictions on proxies
  (i.e.
  strong predictors) of protected information types. 
  Our definition relates proxy use to intermediate computations that
  occur in a program, and identify two essential properties that
  characterize this behavior: \textit{1)} its result is strongly
  associated with the protected information type in question, and
  \textit{2)} it is likely to causally affect the final output of the
  program.
  For a specific instantiation of this definition, we present a
  program analysis technique that detects instances of proxy use in a
  model, and provides a witness that identifies which parts of the
  corresponding program exhibit the behavior.
  Recognizing that not all instances of proxy use of a protected
  information type are inappropriate, we make use of a normative
  judgment oracle that makes this inappropriateness determination for
  a given witness.
  Our repair algorithm uses the witness of an inappropriate proxy use
  to transform the model into one that provably does not exhibit proxy
  use, while avoiding changes that unduly affect classification
  accuracy.
  Using a corpus of social datasets, our evaluation shows that these
  algorithms are able to detect proxy use instances that would be
  difficult to find using existing techniques, and subsequently remove
  them while maintaining acceptable classification performance.
\end{abstract}

\ccsdesc[500]{Security and privacy~Privacy protections}

\keywords{use privacy} %

\maketitle

\section{Introduction}
\label{sect:intro}

Restrictions on information use occupy a central place in privacy
regulations and legal
frameworks~\cite{pipeda,eu16gdpr,teufel08memorandum,s06taxonomy}.
We introduce the term \emph{use privacy} to refer to privacy norms
governing information use.
A number of recent cases have evidenced that inappropriate information
use can lead to violations of both privacy laws~\cite{canada-pc} and
user expectations~\cite{target-nytimes,datta15pets}, prompting calls
for technology to assist with enforcement of use privacy
requirements~\cite{graham-pcast2014}.
In order to meet these regulatory imperatives and user expectations,
companies dedicate resources toward compliance with privacy policies
governing information use~\cite{pasquale15book,graham-pcast2014}.
A large body of work has emerged around use privacy compliance
governing the explicit use of protected information types (see
Tschantz et al.~\cite{tschantz12sp} for a survey).
Such methods are beginning to see deployment in major technology
companies like Microsoft~\cite{sen14sp}.

In this paper, we initiate work on formalizing and enforcing a richer
class of use privacy restrictions---those governing the use of
protected information indirectly through proxies in data-driven
systems.
Data-driven systems include machine learning and artificial
intelligence systems that use large swaths of data about individuals
in order to make decisions about them.
The increasing adoption of these systems in a wide range of sectors,
including advertising, education, healthcare, employment, and credit,
underscores the critical need to address use privacy
concerns~\cite{pasquale15book,graham-pcast2014}.

We start with a set of examples to motivate these privacy concerns and
identify the key research challenges that this paper will tackle to
address them.
In 2012, the department store Target drew flak from privacy advocates
and data subjects for using the shopping history of their customers to
predict their pregnancy status and market baby items based on that
information~\cite{target-nytimes}.
While Target intentionally inferred the pregnancy status and used it
for marketing, the privacy concern persists even if the inference were
not explicitly drawn.
Indeed, the use of health condition-related search terms and browsing
history---\emph{proxies} (i.e., strong predictors) for health
conditions---for targeted advertising have been the basis for legal
action and public concern from a privacy
standpoint~\cite{canada-pc,datta15pets,sunlight}.
Similar privacy concerns have been voiced about the use of personal
information in the Internet of Things
\cite{maddox16dark, hilts16step, nissenbaum16biosensing,
  turow2017aisles}.

\paragraph*{Use privacy}
To address these threats, this paper articulates the problem of
protecting \emph{use privacy} in data-driven systems.

\vspace{2mm}
\noindent
\emph{
Use privacy constraints restrict the use of protected information types and
some of their proxies in data-driven systems.
}
\vspace{2mm}

\paragraph{Setting}
A use privacy constraint may require that health information or its
proxies not be used for advertising.
Indeed there are calls for this form of privacy constraint
\cite{canada-pc,davis16ftc,graham-pcast2014,lipton16privacy}.
In this paper, we consider the setting where a data-driven system is
audited to ensure that it complies with such use privacy constraints.
The auditing could be done by a trusted data processor who is
operating the system or by a regulatory oversight organization who has
access to the data processors' machine learning models and knowledge
of the distribution of the dataset.
In other words, we assume that the data processor does not act to
evade the detection algorithm, and provides accurate information.
This trusted data processor setting is similar to the one assumed in
differential privacy~\cite{dwork06crypto}.

In this setting, it is impossible to guarantee that data processors
with strong background knowledge are not able to infer certain facts
about individuals (e.g., their pregnancy status)~\cite{dwork06icalp}.
Even in practice, data processors often have access to detailed
profiles of individuals and can infer sensitive information about
them~\cite{target-nytimes,turow-daily-you}.
Use privacy instead places a more pragmatic requirement on data-driven
systems: that they simulate ignorance of protected information types
(e.g., pregnancy status) by not using them or their proxies in their
decision-making.
This requirement is met if the systems (e.g., machine learning models)
do not infer protected information types or their proxies (even if
they could) or if such inferences do not affect decisions.

Recognizing that not all instances of proxy use of a protected
information type are inappropriate, our theory of use privacy makes
use of a normative judgment oracle that makes this inappropriateness
determination for a given instance.
For example, while using health information or its proxies for credit
decisions may be deemed inappropriate, an exception could be made for
proxies that are directly relevant to the credit-worthiness of the
individual (e.g., her income and expenses).

\paragraph*{\Proxyuse}
A key technical contribution of this paper is a formalization of
\emph{proxy use} of protected information types in programs.
Our formalization relates proxy use to intermediate computations
obtained by decomposing a program.
We begin with a qualitative definition that identifies two essential
properties of the intermediate computation (the proxy): \textit{1)}
its result perfectly predicts the protected information type in
question, and \textit{2)} it has a causal affect on the final output
of the program.

In practice, this qualitative definition of proxy use is too rigid for
machine learning applications along two dimensions.
First, instead of demanding that proxies are perfect predictors, we
use a standard measure of association strength from the quantitative
information flow security literature to define an
$\epsilon$-\emph{proxy} of a protected information type; here
$\epsilon \in [0,1]$ with higher values indicating a stronger proxy.
Second, qualitative causal effects are not sufficiently informative
for our purpose.
Instead we use a recently introduced causal influence
measure~\cite{qii} to quantitatively characterize influence.
We call it the $\delta$-\emph{influence} of a proxy where
$\delta \in [0,1]$ with higher values indicating stronger influence.
Combining these two notions, we define a notion of \edproxyuse.

We arrive at this program-based definition after a careful examination
of the space of possible definitions.
In particular, we prove that it is impossible for a purely semantic
notion of intermediate computations to support a meaningful notion of
proxy use as characterized by a set of natural properties or axioms
(Theorem~\ref{thm:sem-impossibility}).
The program-based definition arises naturally from this exploration by
replacing semantic decomposition with decompositions of the program.
An important benefit of this choice of restricting the search for
intermediate computations to those that appear in the text of the
program is that it supports natural algorithms for detection and
repair of proxy use.
Our framework is parametric in the choice of a programming language in
which the programs (e.g., machine learnt models) are expressed and the
population to which it is applied.
The choice of the language reflects the level of white-box access that
the analyst has into the program.

\paragraph*{Detection}
We instantiate our definition to a simple programming language that
contains conditionals, arithmetic and logical operations, and
decompositions that involve single variables and associative
arithmetic.
For example, decompositions of linear models include additive sets of
linear terms, and decision forests include subtrees, and sets of
decision trees.
For this instantiation of the definition, we present a program
analysis technique that detects proxy use in a model, and provides a
witness that identifies which parts of the corresponding program
exhibit the behavior (Algorithm~\ref{alg:detect-generic}).
Our algorithm assumes access to the text of a program that computes
the model, as well as a dataset that has been partitioned into
analysis and validation subsets.
The algorithm is program-directed and is directly inspired by the
definition of proxy use.
We prove that the algorithm is complete relative to our instantiation
of the proxy use definition --- it identifies every instance of proxy
use in the program (Theorem~\ref{thm:detect-complete}) and outputs
witnesses (i.e.
intermediate computations that are the proxies).
We provide three optimizations that leverage \emph{sampling},
\emph{pre-computation}, and \emph{reachability} to speed up the
detection algorithm.

\paragraph*{Repair}
If a found instance of proxy use is deemed inappropriate, our repair
algorithm (Algorithm~\ref{alg:repair}) uses the witness to transform
the model into one that provably does not exhibit that instance of
proxy use (Theorem~\ref{thm:repair}), while avoiding changes that
unduly affect classification accuracy.
We leverage the witnesses that localize where in the program a
violation occurs in order to focus repair there.
To repair a violation, we search through expressions local to the
violation, replacing the one which has the least impact on the
accuracy of the model and at the same time reduces the association or
influence of the violation to below the $(\epsilon,\delta)$ threshold.

\paragraph*{Evaluation}

We empirically evaluate our proxy use definition, detection and repair
algorithms on four real datasets used to train decision trees, linear
models, and random forests.
Our evaluation demonstrates the typical workflow for practitioners who
use our tools for a simulated financial services application.
It highlights how they help them uncover more proxy uses than a
baseline procedure that simply eliminates features associated with the
protected information type.
For three other simulated settings on real data sets---contraception
advertising, student assistance, and credit advertising---we find
interesting proxy uses and discuss how the outputs of our detection
tool could aid a normative judgment oracle determine the
appropriateness of proxy uses.
\ifdefined\evenshorterappendix\else We evaluate the performance of the
detection algorithm and show that, in particular cases, the runtime of
our system scales linearly in the size of the model. \fi We demonstrate
the completeness of the detection algorithm by having it discover
artificially injected violations into real data sets.
Finally, we evaluate impact of repair on model accuracy, in
particular, showing a graceful degradation in accuracy as the
influence of the violating proxy increases.

\paragraph*{Closely related work}
The emphasis on restricting use of information by a system rather than
the knowledge possessed by agents distinguishes our work from a large
body of work in privacy (see Smith~\cite{smith15lics} for a survey).
The privacy literature on use restrictions has typically focused on
explicit use of protected information types, and not on proxy use (see
Tschantz et al.~\cite{tschantz12sp} for a survey and Lipton and
Regan~\cite{lipton16privacy}).
Recent work on discovering personal data use by black-box web services
focuses mostly on explicit use of protected information types by
examining causal effects \cite{barford14www, datta15pets,
  englehardt14man, guha10, hannak13www, hannak14imc, xray, sunlight,
  liu16ndss, wills12wpes, vissers14hotpets}); some of this work also
examines associational effects~\cite{xray, sunlight}.
Associational effects capture some forms of proxy use but not others
as we argue in Section~\ref{sec:defn}.

In a setting similar to ours of a trusted data processor, differential
privacy~\cite{dwork06crypto} protects against a different type of
privacy harm.
For a computation involving data contributed by a set of individuals,
differential privacy minimizes any knowledge gains by an adversary
that are caused by the contribution of a single individual.
This requirement, however, says nothing about what information types
about an individual are actually used by the data processor, the
central concern of use privacy.

Lipton and Regan's notion of ``effectively private" captures the idea
that a protected feature is not explicitly used to make decisions, but
does not account for proxy use~\cite{lipton16privacy}.
Prior work on fairness has also recognized the importance of dealing
with proxies in machine learning systems~\cite{dwork12itcs,
  feldman15kdd, fairtest}.
However treatments of proxy use considered there do not match the
requirements of use privacy.
We elaborate on this point in Section~\ref{sec:defn}.
In Section~\ref{sect:related}, we provide a more detailed comparison
with related work highlighting that use privacy enhancing technology
(PET) complements existing work on PETs.
It is not meant to supplant other PETs geared toward restricting data
collection and release.
While the results of this paper represent significant progress toward
enabling use privacy, as elaborated in Section~\ref{sec:discussion}, a
host of challenging problems remain open.

\paragraph*{Contributions}

In summary, we make the following contributions:

\begin{itemize}
\item An articulation of the problem of protecting \emph{use privacy}
  in data-driven systems.
  Use privacy restricts the use of protected information types and
  some of their proxies (i.e., strong predictors) in automated
  decision-making systems (\S\ref{sect:intro}, \ref{sec:use-privacy}).
\item A formal definition of \emph{proxy use}---a key building block
  for use privacy--and an axiomatic basis for this definition
  (\S\ref{sec:defn}).
\item An algorithm for detection and tracing of an instantiation of
  proxy use in a machine learnt program, and proof that this algorithm
  is sound and complete (\S\ref{sect:detection}).
\item A repair algorithm that provably removes violations of the proxy
  use instantiation in a machine learning model that are identified by
  our detection algorithm and deemed inappropriate by a normative
  judgment oracle (\S\ref{sect:repair}).
\item An implementation and evaluation of our
  approach on popular machine learning algorithms
  applied to real datasets (\S\ref{sect:eval}).
\end{itemize}

\ifdefined\arXiv\else An extended version\cite{datta2017use} of this
paper includes additional evaluation and further details on the
datasets employed.\fi

\cut{
  \paragraph*{Examples of proxy use}
\begin{itemize}
\item Indonesian Contraception Dataset: Education is a proxy for
  religion
\item Census Income Prediction Dataset: Age is a proxy for marital
  status.
  Can be deemed permitted.
  Education does not have proxy use.
\item Student Loan Dataset: Disability is a sensitive attribute (not
  sure what the findings here are)
  \item Census Dataset: Shayak working on this.
  \item Portuguese Alcohol Consumption Dataset Student performance: ??
    is working on this.
  \end{itemize}
}

\section{Use Privacy}
\label{sec:use-privacy}

We use the Target example described earlier in the paper to motivate 
our notion of use privacy. Historically, data collected in a context of interaction 
between a retailer and a consumer is not expected to result in flows of 
health information. However, such flow constraints considered in 
significant theories of privacy (e.g., see Nissenbaum \cite{nissenbaum09book}) 
cannot be enforced because of possible statistical
inferences. In particular, prohibited information types (e.g., pregnancy
status) could be inferred from legitimate flows (e.g., shopping history). Thus, 
the theory of use privacy instead ensures that the
data processing systems ``simulate ignorance'' of protected information 
types (e.g., pregnancy status) and their proxies (e.g., purchase history) 
by not using them in their decision-making. 
Because not all instances of proxy use of a protected information type
are inappropriate, our theory of use privacy makes use of a normative judgment
oracle that makes this inappropriateness determination for a given instance.

We model the 
personal data processing system as a program $p$. The use privacy constraint 
governs a protected information type $Z$. Our definition of use privacy 
makes use of two building blocks: (1) a function that given $p$, $Z$, and a 
population distribution $\population$ returns a witness $w$ of proxy use of $Z$ 
in a program $p$ (if it exists); and (2) a normative judgment oracle 
$\oracle(w)$ that given a specific witness returns a judgment on whether the 
specific proxy use is appropriate ({\sc true}) or not ({\sc false}). 

\begin{defn}[Use Privacy] Given a program $p$, protected information type $Z$,
  normative judgment oracle $\oracle$, and population distribution $\population$, use privacy
  in a program $p$ is violated if there exists a witness $w$ in $p$ of \proxyuse of $Z$
  in $\population$ such that $\oracle(w)$ returns {\sc false}.
\end{defn}

In this paper, we formalize the computational component of the above definition  of
use privacy, by formalizing what it means for an algorithm to use a protected
information type directly or through proxies (\S\ref{sec:defn}) and designing 
an algorithm to detect proxy uses in programs (\S\ref{sect:detection}). We assume that 
the normative judgment oracle is given to us and use it to identify inappropriate 
proxy uses and then repair them (\S\ref{sect:repair}). 
In our experiments, we illustrate how such an oracle would use the outputs of our
proxy use analysis and recommend the repair of uses deemed inappropriate by it (\S\ref{sect:eval}).
 
This definition cleanly separates computational considerations 
that are automatically enforceable and ethical judgments that require input from 
human experts. This form of separation exists also in some prior work on 
privacy \cite{garg-jia-datta-ccs-2011} and fairness \cite{Dwork12}.

\section{Proxy Use: A Formal Definition}
\label{sec:defn}

We now present an axiomatically justified, formal definition of proxy use in
data-driven programs. Our definition for proxy use of a
protected information type involves \emph{decomposing} a program to find an
intermediate computation whose result exhibits two properties:
\begin{itemize}
  \item \emph{Proxy}: strong association with the protected type
  \item \emph{Use}: causal influence on the output of the program
\end{itemize}

In \S~\ref{sec:defn:examples}, we present a sequence of examples to illustrate
the challenge in identifying proxy use in systems that operate on data
associated with a protected information type. In doing so, we will also
contrast our work with closely-related work in privacy and fairness.
In \S\ref{sec:notation}, we formalize the notions of proxy
and use, preliminaries to the definition. The definition itself is presented
in \S\ref{sec:definition} and \S\ref{sec:defn:quant}. Finally, in
\S\ref{sec:defn:impossibility}, we provide an axiomatic characterization of the
notion of proxy use that guides our definitional choices.
\cut{In particular, we
prove an important semantic impossibility result that steers us towards the
syntactic definition presented in \S\ref{sec:definition} and
\S\ref{sec:defn:quant}.} We note that readers keen to get to the detection and
repair mechanisms may skip \S\ref{sec:defn:impossibility} without loss of
continuity.

\subsection{Examples of Proxy Use}
\label{sec:defn:examples}

Prior work on detecting use of protected information
types~\cite{datta2015opacity,sunlight,fairtest,feldman15disparate} and
leveraging knowledge of detection to eliminate inappropriate
uses~\cite{feldman15disparate} have treated the system as a black-box.
Detection relied either on experimental access to the
black-box~\cite{datta2015opacity,sunlight} or observational data about
its behavior~\cite{fairtest,feldman15disparate}.
Using a series of examples motivated by the Target case, we motivate
the need to peek inside the black-box to detect proxy use.

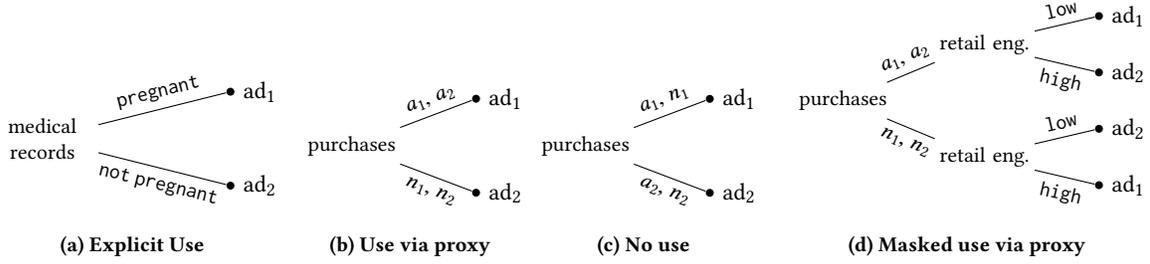
\begin{figure*}[t]
\centering

\tikzstyle{bag}=[text width=4em,text centered,font=\small]
\tikzstyle{end}=[circle,minimum width=3pt,fill,inner sep=0pt]

\tikzstyle{level 1}=[level distance=2.5cm, sibling distance=1.25cm]
\tikzstyle{level 2}=[level distance=2.5cm, sibling distance=0.75cm]
\subcaptionbox{Explicit Use\label{fig:tree-explicit}}{
  \begin{tikzpicture}[grow=right,sloped]
    \node[bag] {medical records}
    child {
      node[end, label=right:{ad$_2$}] {}
      edge from parent
      node[below] {\small$\mathtt{not\ pregnant}$}
    }
    child {
      node[end, label=right:{ad$_1$}] {}
      edge from parent
      node[above] {\small$\mathtt{pregnant}$}
    }
    ;
  \end{tikzpicture}
}\hspace{-0.5em}%
\tikzstyle{level 1}=[level distance=1.65cm, sibling distance=1.25cm]
\tikzstyle{level 2}=[level distance=1.65cm, sibling distance=0.75cm]
\subcaptionbox{Use via proxy\label{fig:tree-redline}}{
  \begin{tikzpicture}[grow=right,sloped]
    \node[bag] {purchases}
    child {
      node[end, label=right:{ad$_2$}] {}
      edge from parent
      node[below] {\small$n_1,n_2$}
    }
    child {
      node[end, label=right:{ad$_1$}] {}
      edge from parent
      node[above] {\small$a_1,a_2$}
    }
    ;
  \end{tikzpicture}
}\hspace{-0.5em}%
\subcaptionbox{No use\label{fig:tree-benign}}{
  \begin{tikzpicture}[grow=right,sloped]
    \node[bag] {purchases}
    child {
      node[end, label=right:{ad$_2$}] {}
      edge from parent
      node[below] {\small$a_2,n_2$}
    }
    child {
      node[end, label=right:{ad$_1$}] {}
      edge from parent
      node[above] {\small$a_1,n_1$}
    }
    ;
  \end{tikzpicture}
}\hspace{0em}%
\tikzstyle{level 1}=[level distance=1.8cm, sibling distance=1.5cm]
\tikzstyle{level 2}=[level distance=1.6cm, sibling distance=0.75cm]
\subcaptionbox{Masked use via proxy\label{fig:tree-masked}}{
  \begin{tikzpicture}[grow=right,sloped]
    \node[bag] {purchases}
    child {
      node[bag,align=right] {retail eng.}
      child {
        node[end,label=right:{ad$_1$}] {}
        edge from parent
        node[below] {\small$\mathtt{high}$}
      }
      child {
        node[end,label=right:{ad$_2$}] {}
        edge from parent
        node[above] {\small$\mathtt{low}$}
      }
      edge from parent
      node[below] {\small$n_1,n_2$}
    }
    child{
      node[bag,align=right] {retail eng.}
      child {
        node[end,label=right:{ad$_2$}] {}
        edge from parent
        node[below] {\small$\mathtt{high}$}
      }
      child {
        node[end,label=right:{ad$_1$}] {}
        edge from parent
        node[above] {\small$\mathtt{low}$}
      }
      edge from parent
      node[above] {\small$a_1,a_2$}
    };
  \end{tikzpicture}
}

 \caption{Examples of models (decision trees) used by a retailer
  for offering medicines and for selecting advertisements to show to
  customers.
  The retailer uses pregnancy status, past purchases, and customer's
  level of retail engagement.
  Products $a_1$ and $a_2$ are associated with
  pregnancy (e.g., prenatal vitamins, scent-free lotions) whereas
  products  $n_1$ and $n_2$ are associated with a lack of
  pregnancy (e.g., alcohol, camping gear); all four products are
  equally likely.
  Retail engagement, (\texttt{high} or \texttt{low}), indicating whether the customer views ads or not, is
independent of pregnancy.}
\label{fig:proxy-usage-examples}
\end{figure*}

\begin{example}
  \label{ex:proxy1}
  (Explicit use, Fig.~\ref{fig:tree-explicit}) A retailer explicitly uses pregnancy status from prescription data available at its pharmacy to market baby products. \cut{Figure~\ref{fig:tree-explicit} shows such an algorithm.}
\end{example}

This form of explicit use of a protected information type
can be discovered by existing black-box experimentation methods that establish
causal effects between inputs and outputs (e.g., see
\cite{datta2015opacity,sunlight}).

\begin{example}
  \label{ex:proxy2}
(Inferred use, Fig.~\ref{fig:tree-redline})
Consider a situation where purchase history can be used to accurately predict pregnancy status.
A retailer markets specific products to individuals who have recently purchased products
indicative of pregnancy (e.g., $a_1,a_2 \in \text{purchases}$). \cut{Figure ~\ref{fig:tree-redline} is an example of such an algorithm.}
\end{example}

This example, while very similar in effect, does not use health information
directly. Instead, it infers pregnancy status via associations and then uses
it.  Existing methods (see \cite{feldman15disparate,fairtest}) can detect such
associations between protected information types and outcomes in observational
data.

\begin{example}
  \label{ex:proxy3}
  (No use, Fig.~\ref{fig:tree-benign}) Retailer uses some uncorrelated selection of products ($a_1,n_1 \in \text{purchases}$) to suggest ads. \cut{Figure ~\ref{fig:tree-benign} is an example of such an algorithm.}
\end{example}

In this example, even though the retailer could have inferred pregnancy status from the purchase history, no such inference was used in marketing products. As associations are commonplace, a definition of use disallowing such benign use of associated data would be too restrictive for practical enforcement.

\begin{example}
  \label{ex:proxy4}
  (Masked proxy use, Fig.~\ref{fig:tree-masked})
  Consider a more insidious version of Example~\ref{ex:proxy2}. To mask the
association between the outcome and pregnancy status, the company also markets baby products to people
who are not pregnant, but have low retail engagement, so these advertisements would not be viewed in any case. \cut{Figure~\ref{fig:tree-masked} is an example of such an algorithm.}
\end{example}

While there is no association between pregnancy and outcome in both
Example~\ref{ex:proxy3} and Example~\ref{ex:proxy4}, there is a key difference
between them.  In Example~\ref{ex:proxy4}, there is an intermediate computation
based on aspects of purchase history that is a predictor for pregnancy status,
and this predictor is used to make the decision, and therefore is a case of
proxy use. In contrast, in Example~\ref{ex:proxy3}, the intermediate
computation based on purchase history is uncorrelated with pregnancy status.
Distinguishing between these examples by measuring associations using black box
techniques is non-trivial. Instead, we leverage white-box access to the code of
the classifier to identify the intermediate computation that serves as a proxy for
pregnancy status. Precisely identifying the particular proxy used also aids the
normative decision of whether the proxy use is appropriate in this setting.

\subsection{Notation and Preliminaries}
\label{sec:notation}

We assume individuals are drawn from a population distribution $\population$,
in which our definitions are parametric.
\cut{Our definitions are parametric in the choice of the population $\population$,
and identifies proxy use for the chosen population.}
Random variables $W, X, Y, Z, \ldots$ are functions over $\population$, and the
notation $W \in \cal W$ represents that the type of random variable is $W :
\cal P \to \cal W$.  An important random variable used throughout the paper
is $\features$, which represents the vector of features of an individual that
is provided to a predictive model. A predictive model is denoted by
$\pairpop{\features}{\model}$, where $\model$ is a function that operates on
$\features$. For simplicity, we assume that
$\population$ is discrete, and that models are deterministic. Table~\ref{tab:notation}
summarizes all the notation used in this paper, in addition to the notation for
programs that is introduced later in the paper.

\setlength{\belowcaptionskip}{0pt}

\begin{table}
  \centering
  \begin{tabular}{|r | p{0.75\columnwidth} |}
    \hline
    $f$             & A function \\
    $\pairpop{\features}{\model}$ & A model, which is a function $\model$ used for prediction, operating on random variables $\features$, in population $\population$\\
               $X$             & A random variable\\
               $p$             & A program \\
    $\pairpop{\features}{p}$            & A syntactic model, which is a program $p$, operating on random variables $\features$\\
               $\subst{p_1}{X}{p_2}$ & A substitution of $p_1$ in place of $X$ in $
    p_2 $\\
    $\features$         & A sequence of random variables \\
    \hline
\end{tabular}
\caption{Summary of notation used in the paper
\label{tab:notation}
}
\end{table}

\setlength{\belowcaptionskip}{-10pt}
\subsubsection{Proxies}
\label{sec:defn:proxy}

A \emph{perfect proxy} for a random variable $Z$ is a random variable $X$ that
is perfectly correlated with $Z$.  Informally, if $X$ is a proxy of $Z$, then
$X$ or $Z$ can be interchangeably used in any computation over the same
distribution. One way to state this is to require that $\Pr(X = Z) = 1$, i.e.
$X$ and $Z$ are equal on the distribution. However, we require our definition
of proxy to be invariant under renaming. For example, if $X$ is $0$ whenever
$Z$ is $1$ and vice versa, we should still identify $X$ to be a proxy for $Z$.
In order to achieve invariance under renaming, our definition only requires the
existence of mappings between $X$ and $Z$, instead of equality.

\begin{defn}[Perfect Proxy]
\label{def:perfect-proxy}
A random variable $X \in \cal X$ is a \emph{perfect proxy for
  $Z \in \cal Z$} if there exist functions
$f : {\cal X} \to {\cal Z}, g : \cal{Z} \to \cal{X}$, such that
$\Pr(Z = f(X)) = \Pr(g(Z) = X) = 1$.
\end{defn}

While this notion of a proxy is too strong in practice, it is useful
as a starting point to explain the key ideas in our definition of
proxy use.
This definition captures two key properties of proxies,
\emph{equivalence} and \emph{invariance under renaming}.

\paragraph{Equivalence} Definition~\ref{def:perfect-proxy} captures
the property that proxies admit predictors in both directions: it is
possible to construct a predictor of $X$ from $Z$, and vice versa.
This condition is required to ensure that our definition of proxy only
identifies the part of the input that corresponds to the protected
attribute and not the input attribute as a whole.
For example, if only the final digit of a zip code is a proxy for
race, the entirety of the zip code will not be identified as a proxy
even though it admits a predictor in one direction.
Only if the final digit is used, that use will be identified as proxy
use.

The equivalence criterion distinguishes benign use of
associated information from proxy use as illustrated in the next
example.
For machine learning in particular, this is an important pragmatic requirement; given enough input
features one can expect any protected class to be predictable from the
set of inputs.
In such cases, the input features taken together are a strong
associate in one direction, and prohibiting such one-sided associates from being
used would rule out most machine learnt models.

\begin{example}
  Recall that in Figure~\ref{fig:proxy-usage-examples}, $a_1, a_2$ is
  a proxy for pregnancy status.
  In contrast, consider Example~\ref{ex:proxy3}, where purchase
  history is an influential input to the program that serves ads.
  Suppose that the criteria is to serve ads to those with
  $\mathit{a_1,n_1}$ in their purchase history.
  According to Definition~\ref{def:perfect-proxy}, neither purchase
  history or $a_1,n_1$ are proxies, because pregnancy status does not
  predict purchase history or $a_1,n_1$.
  However, if Definition~\ref{def:perfect-proxy} were to allow
  one-sided associations, then purchase history would be a proxy
  because it can predict pregnancy status.
  This would have the unfortunate effect of implying that the benign
  application in Example~\ref{ex:proxy3} has proxy use of pregnancy
  status.
\end{example}

\paragraph{Invariance under renaming} This definition of a proxy is
invariant under renaming of the values of a proxy.
Suppose that a random variable evaluates to $1$ when the protected
information type is $0$ and vice versa, then this definition still
identifies the random variable as a proxy.

\subsubsection{Influence}
\label{sec:defn:influence}

Our definition of influence aims to capture the presence of a causal
dependence between a variable and the output of a function.
Intuitively, a variable $x$ is influential on $f$ if it is possible to
change the value of $f$ by changing $x$ while keeping the other input
variables fixed.

\begin{defn}
  For a function $f(x, y)$, $x$ is influential if and only if there
  exists values $x_1$, $x_2$, $y$, such that
  $f(x_1, y) \not= f(x_2, y)$.
\end{defn}

In Figure~\ref{fig:tree-explicit}, pregnancy status is an influential
input of the system, as just changing pregnancy status while keeping
all other inputs fixed changes the prediction.
Influence, as defined here, is identical to the notion of interference
used in the information flow literature.%

\subsection{Definition}
\label{sec:definition}

We use an abstract framework of program syntax to reason about
programs without specifying a particular language to ensure that our
definition remains general.
Our definition relies on syntax to reason about decompositions of
programs into intermediate computations, which can then be identified
as instances of proxy use using the concepts described above.

\paragraph{Program decomposition}
We assume that models are represented by programs. For a set of random
variables $\features$, $\pairpop{\features}{p}$ denotes the assumption that
$p$ will run on the variables in $\features$. Programs are given meaning by a
denotation function $\denote{\cdot}_\features$ that maps programs to
functions. If $\pairpop{\features}{p}$, then $\denote{p}$ is a function on
variables in $\features$, and $\denote{p}(\features)$ represents the random
variable of the outcome of $p$, when evaluated on the input random variables
$\features$. Programs support substitution of free variables with other
programs, denoted by $\subst{p_1}{X}{p_2}$, such that if $p_1$ and $p_2$
programs that run on the variables $\features$ and $\features, X$,
respectively, then $\subst{p_1}{X}{p_2}$ is a program that operates on
$\features$.

A decomposition of program $p$ is a way of rewriting $p$ as two programs $p_1$
and $p_2$ that can be combined via substitution to yield the original program.

\begin{defn}[Decomposition]
  \label{def:decomposition}
  Given a program $p$, a decomposition $(p_1, X, p_2)$
  consists of two programs $p_1$, $p_2$, and a fresh
  variable $X$, such that $p = \subst{p_1}{X}{p_2}$.
\end{defn}

For the purposes of our proxy use definition we view the first
component $p_1$ as the intermediate computation suspected of proxy use, and
$p_2$ as the rest of the computation that takes in $p_1$ as an input.

\begin{defn}[Influential Decomposition]
  Given a program $p$, a decomposition $(p_1, X, p_2)$ is influential
  iff $X$ is influential in $p_2$.
\end{defn}

\paragraph{Main definition}

\begin{defn}[Proxy Use]
\label{def:proxy-usage}
A program $\pairpop{\features}{p}$ has proxy use of $Z$ if there
exists an influential decomposition $(p_1, X, p_2)$ of
$\pairpop{\features}{p}$, and $\denote{p_1}(\features)$ is a proxy for
$Z$.
\end{defn}

\begin{example}
  In Figure~\ref{fig:tree-masked}, this definition would identify proxy use using the decomposition $(p_1, U, p_2)$, where $p_2$ is the entire tree, but with the condition  $(a_1,a_2 \in \text{purchases})$ replaced by the variable $U$. In this example, $U$ is influential in $p_2$, since changing the value of $U$ changes the outcome. Also, we assumed that the condition $(a_1,a_2 \in \text{purchases})$ is a perfect predictor for pregnancy, and is therefore a proxy for pregnancy. Therefore, according to our definition of proxy use, the model in \ref{fig:tree-masked} has proxy use of pregnancy status.
\end{example}

\subsection{A Quantitative Relaxation}
\label{sec:defn:quant}

Definition~\ref{def:proxy-usage} is too strong in one sense and too
weak in another.
It requires that intermediate computations be perfectly correlated
with a protected attribute, and that there exists \emph{some} input,
however improbable, in which the result of the intermediate
computation is relevant to the model.
For practical purposes, we would like to capture imperfect proxies
that are strongly associated with an attribute, but only those whose
influence on the final model is appreciable.
To relax the requirement of perfect proxies and non-zero influence, we
quantify these two notions to provide a parameterized definition.
Recognizing that neither perfect privacy nor perfect utility are
practical, the quantitative definition provides a means for navigating
privacy vs.
utility tradeoffs.

\paragraph{$\epsilon$-proxies}

We wish to measure how strongly a random variable $X$ is a proxy for a
random variable $Z$.
Recall the two key requirements from the earlier definition of a
proxy: (i) the association needs to be capture equivalence and measure association in both directions, and (ii) the
association needs to be invariant under renaming of the random
variables.
The \emph{variation of information metric}
$d_{\text{var}}(X, Z) = H(X | Z) + H(Z | X)$~\cite{cover2012elements}
is one measure that satisfies these two requirements.
The first component in the metric, the conditional entropy of $X$
given $Z$, $H(X|Z)$, measures how well $X$ can be predicted from $Z$,
and $H(Z|X)$ measures how well $Z$ can be predicted from $X$, thus
satisfying the requirement for the metric measuring association in both directions.
Additionally, one can show that conditional entropies are invariant
under renaming, thus satisfying our second criteria.
To obtain a normalized measure in $[0, 1]$, we choose
$1 - \frac{d_{\text{var}}(X, Z)}{H(X, Z)}$ as our measure of
association, where the measure being $1$ implies perfect proxies, and
$0$ implies statistical independence.
Interestingly, this measure is identical to normalized mutual
information~\cite{cover2012elements}, a standard measure that has also been
used in prior work in identifying associations in outcomes of machine learning
models~\cite{fairtest}.

\begin{defn}[Proxy Association]
  Given two random variables $X$ and $Z$, the strength of a proxy is
  given by normalized mutual information,
\[
  d(X, Z) \defeq 1 - \frac{H(X | Z) +
  H(Z | X)}{H(X, Z)}
\]
where $X$ is defined to be an $\epsilon$-proxy for $Z$ if
$d(X, Z) \geq \epsilon$.
\end{defn}

We do not present the complexity of association computation
independently of detection as we rely on pre-computations to reduce
the amortized runtime of the entire detection algorithm.
The complexity as part of our detection algorithm is discussed in
Appendix~\ref{sec:estimates}.

\paragraph{$\delta$-influential decomposition}

Recall that for a decomposition $(p_1, X, p_2)$, in the qualitative
sense, influence is interference which implies that there exists $x$,
$x_1$, $x_2$, such that
$\denote{p_2}(x, x_1) \not= \denote{p_2}(x, x_2)$.
Here $x_1$, $x_2$ are values of $p_1$, that for a given $x$, change
the outcome of $p_2$.
However, this definition is too strong as it requires only a single
pair of values $x_1$, $x_2$ to show that the outcome can be changed by
$p_1$ alone.
To measure influence, we quantify interference by using Quantitative
Input Influence (QII), a causal measure of input influence introduced
in \cite{qii}.
In our context, for a decomposition $(p_1, X, p_2)$, the influence of
$p_1$ on $p_2$ is given by:
\begin{equation*}
  \iota(p_1, p_2) \defeq \mathbb{E}_{\features,\features' \samp
    \population}\prob{\denote{p_2}(\features, \denote{p_1}(\features))
    \not= \denote{p_2}(\features, \denote{p_1}(\features'))}.
\end{equation*}
Intuitively, this quantity measures the likelihood of finding randomly
chosen values of the output of $p_1$ that would change the outcome of
$p_2$.
Note that this general definition allows for probabilistic models
though in this work we only evaluate our methods on deterministic
models.

The time complexity of influence computation as part of our detection
algorithm can be found in Appendix~\ref{sec:estimates}, along with
discussion on estimating influence.

\begin{defn}[Decomposition Influence]
  Given a decomposition $(p_1, X, p_2)$, the influence of the
  decomposition is given by the QII of $X$ on $p_2$.
  A decomposition $(p_1, X, p_2)$ is defined to be
  $\delta$-influential if $\iota(p_1, p_2) > \delta$.
\end{defn}

\paragraph{$(\epsilon,\delta)$-proxy use}

Now that we have quantitative versions of the primitives used in
Definition~\ref{def:proxy-usage}, we are in a position to define quantitative
proxy use (Definition~\ref{def:quant-usage}). The structure of this
definition is the same as before, with quantitative measures substituted in for
the qualitative assertions used in Definition~\ref{def:proxy-usage}.

\begin{defn}[$(\epsilon,\delta)$-proxy use]
\label{def:quant-usage}
A program $\pairpop{\features}{p}$  has
$(\epsilon,\delta)$-proxy use of random variable $Z$ iff there
exists a $\delta$-influential decomposition $(p_1, X, p_2)$, such that
$\denote{p}(\features)$ is an $\epsilon$-proxy for $Z$.
\end{defn}

This definition is a strict relaxation of
Definition~\ref{def:proxy-usage}, which reduces to $(1, 0)$-proxy use.

\subsection{Axiomatic Basis for Definition}
\label{sec:defn:impossibility}

We now motivate our definitional choices by reasoning about a natural set of
properties that a notion of proxy use should satisfy.  We first prove an
important impossibility result that shows that no definition of proxy use can
satisfy four natural semantic properties of proxy use.  The central reason behind the
impossibility result is that under a purely semantic notion of function
composition, the causal effect of a proxy can be made to disappear.  Therefore,
we choose a syntactic notion of function composition for the definition of
proxy use presented above. The syntactic definition of proxy use is
characterized by syntactic properties which map very closely to the semantic
properties.

\begin{property}
\label{axm:explicit-sem}
\emph{(Explicit Use)}
If $Z$ is an influential input of the model $\pairpopbrace{\features,Z}{\model}$, then
$\pairpopbrace{\features,Z}{\model}$ has proxy use of $Z$.
\end{property}
This property identifies the simplest case of proxy use: if an input to
the model is influential, then the model exhibits proxy use of that input.

\begin{property}
\label{axm:prep-sem}
\emph{(Preprocessing)} If a model $\pairpopbrace{\features,X}{\model}$
has proxy use of random variable $Z$, then for any function $f$ such
that $\prob{f(\features) = X} = 1$, let
$\model'(x) \defeq \model(x, f(x))$.
Then, $\pairpop{\features}{\model'}$ has proxy use of $Z$.
\end{property}
This property covers the essence of proxy use where instead of being provided a
protected information type explicitly, the program uses a strong predictor for it
instead. This property states that models that use inputs explicitly
and via proxies should not be differentiated under a reasonable theory of proxy use.

\begin{property}
\label{axm:dummy-sem}
\emph{(Dummy)} Given $\pairpop{\features}{\model}$, define $\model'$
such that for all $x, x'$, $\model'(x, x') \defeq \model(x)$, then
$\pairpop{\features}{\model}$ has proxy use for some $Z$ iff
$\pairpop{\{\features, X\}}{\model'}$ has proxy use of $Z$.
\end{property}
This property states that the addition of an input to a model that is
not influential, i.e., has no effect on the outcomes of the model, has
no bearing on whether a program has proxy use or not.
This property is an important sanity check that ensures that models
aren't implicated by the inclusion of inputs that they do not use.

\begin{property}
\label{axm:indep-sem}
\emph{(Independence)}
If $\features$ is independent of $Z$ in $\population$, then $\pairpop{\features}{\model}$ does not have
proxy use of $Z$.
\end{property}
Independence between the protected information type and the inputs ensures that the model
cannot infer the protected information type for the population $\population$.  This
property captures the intuition that if the model cannot infer the protected information type
then it cannot possibly use it.

While all of these properties seem intuitively desirable, it turns out
that these properties can not be achieved simultaneously.

\begin{thm}\label{thm:sem-impossibility}\label{THM:SEM-IMPOSSIBILITY} %
  No definition of proxy use can satisfy Properties
  \ref{axm:explicit-sem}-\ref{axm:indep-sem} simultaneously.
\end{thm}

See Appendix~\ref{app:impossibility-proof} for a proof of the impossibility result
and a discussion.
The key intuition behind this result is that Property \ref{axm:prep-sem} requires proxy
use to be preserved when an input is replaced with a function that
predicts that input via composition. However, with a purely semantic notion of function
composition, after replacement, the proxy may get canceled out.
To overcome this impossibility
result, we choose a more syntactic notion of function composition, which is
tied to how the function is represented as a program, and looks for evidence
of proxy use within the representation.

We now proceed to the axiomatic justification of our definition of proxy use.
As in our attempt to formalize a semantic definition, we base our definition on
a set of natural properties given below. These are
syntactic versions of their semantic counterparts defined earlier.

\begin{property}
\label{axm:syn-explicit}
\emph{(Syntactic Explicit Use)}
If $X$ is a proxy of $Z$, and $X$ is an
influential input of $\pairpopbrace{\features, X}{p}$, then $\pairpopbrace{\features, X}{p}$ has proxy use.
\end{property}

\begin{property}
\label{axm:syn-preprocessing}
\emph{(Syntactic Preprocessing)}
If $\pairpopbrace{\features, X}{p_1}$ has proxy use of $Z$, then for
any $p_2$ such that $\prob{\denote{p_2}(\features) = X} = 1$,
$\pairpop{\features}{\subst{p_2}{X}{p_1}}$ has proxy use of $Z$.
\end{property}

\begin{property}
\label{axm:syn-dummy}
\emph{(Syntactic Dummy)}
Given a program $\pairpop{\features}{p}$,
$\pairpop{\features}{p}$ has proxy use for some $Z$ iff $\pairpopbrace{\features, X}{p}$ has proxy use of $Z$.
\end{property}

\begin{property}
\label{axm:syn-independence}
\emph{(Syntactic Independence)}
If $\features$ is independent of $Z$, then $\pairpop{\features}{p}$ does not have
proxy use of $Z$.
\end{property}

Properties \ref{axm:syn-explicit} and \ref{axm:syn-preprocessing}
together characterize a complete inductive definition, where the
induction is over the structure of the program.
Suppose we can decompose programs $p$ into $(p_1, X, p_2)$ such that
$p = \subst{p_1}{X}{p_2}$.
Now if $X$, which is the output of $p_1$, is a proxy for $Z$ and is
influential in $p_2$, then by Property~\ref{axm:syn-explicit}, $p_2$
has proxy use.
Further, since $p = \subst{p_1}{X}{p_2}$, by
Property~\ref{axm:syn-preprocessing}, $p$ has proxy use.
This inductive definition where we use Property~\ref{axm:syn-explicit}
as the base case and Property~\ref{axm:syn-preprocessing} for the
induction step, precisely characterizes
Definition~\ref{def:proxy-usage}.
Additionally, it can be shown that Definition~\ref{def:proxy-usage}
also satisfies Properties~\ref{axm:syn-dummy}
and~\ref{axm:syn-independence}.
Essentially, by relaxing our notion of function composition to a
syntactic one, we obtain a practical definition of proxy use
characterized by the natural axioms above.

\section{Detecting \ProxyUse}
\label{sect:detection}

\cut{
\pxm{Insights:} Our detection algorithm enumerates the sub-expressions
in a model in order to test each for \edproxyuse.
Association computations are cheaper for each sub-expression and are
done first while the influence computation is done only if the
$ \epsilon $ threshold is satisfied.
More speedup is achieved the amortized pre-computing the distribution
of inputs to and outputs from each subexpression which can then be
looked up instead of having to re-evaluate large parts of the model
for each subexpression.
Influence computations are further sped-up by ignoring parts of
datasets that do not reach the expression for which influence is
measured.
Models that heavily employ the if-then-else construct (such as
decision trees) greatly benefit from this optimization as the
proportion of the dataset that reach the bodies of the if-then-else
tend to diminish exponentially with depth (or by approximately half
each time a condition is checked).
\pxm{The final thing is not implemented:} For decision trees, another
optimization can be employed which makes the influence computation
linear in the size dataset instead of quadratic.}

In this section, we present an algorithm for identifying \proxyuse of
specified variables in a given machine-learning model
(Algorithm~\ref{alg:detect-generic-informal},
Appendix~\ref{app:detection} contains a more formal presentation of
the algorithm for the interested reader).
The algorithm is program-directed and is directly inspired by the
definition of \proxyuse in the previous section.
We prove that the algorithm is complete in a strong sense --- it
identifies every instance of \proxyuse in the program
(Theorem~\ref{thm:detect-complete}).
We also describe three optimizations that speed up the detection
algorithm: sampling, reachability analysis, and contingency tables.

\begin{algorithm}[t]
  \caption{Detection for expression programs.
    \label{alg:detect-generic-informal}}
  \begin{algorithmic}
    \Require association (\assocmeasure{}), influence(\infmeasure{}) measures
    \Procedure{ProxyDetect}{$p, \features, Z, \epsilon, \delta$}
    \State $P \gets \varnothing$
    \For{each subprogram  $p_1$ appearing in $p$}
      \For{each program $p_2$ such that $[p_2/u]p_1 = p$}
        \If{$\iota(p_1, p_2) \geq \delta \land d(\llbracket p_1 \rrbracket(\features), Z) \geq \epsilon$}
          \State $P \gets P \cup \{(p_1, p_2)\}$
        \EndIf
      \EndFor
    \EndFor
    \State \Return $P$
    \EndProcedure
\end{algorithmic}
\end{algorithm}

\subsection{Environment Model}
\label{sec:detecting:model}

The environment in which our detection algorithm operates is comprised
of a data processor, a dataset that has been partitioned into analysis
and validation subsets, and a machine learning model trained over the
analysis subset.
We assume that the data processor does not act to evade the detection
algorithm, and the datasets correspond to a representative sample from the
population we wish to test proxy use with respect to.
Additionally, we assume that information types we wish to detect
proxies of are also part of the validation data.
We discuss these points further in Section~\ref{sec:discussion}.

For the rest of this paper we focus on an instance of the proxy use
definition, where we assume that programs are written in the simple
expression language shown in Figure~\ref{fig:language}.
However, our techniques are not tied to this particular language, and
the key ideas behind them apply generally.
This language is rich enough to support commonly-used models such as
decision trees, linear and logistic regression, Naive Bayes, and
Bayesian rule lists.
Programs are functions that evaluate arithmetic terms, which are
constructed from real numbers, variables, common arithmetic
operations, and if-then-else ($\mathbf{ite}(\cdot, \cdot, \cdot)$)
terms.
Boolean terms, which are used as conditions in $\mathbf{ite}$ terms,
are constructed from the usual connectives and relational operations.
Finally, we use $\lambda$-notation for functions, i.e., $\lambda x .
e$ denotes a function over $x$ which evaluates $e$ after replacing all
instances of $x$ with its argument.
Details on how machine learning models such as linear models, decision
trees, and random forests are translated to this expression language
are discussed in Appendix~\ref{app:translation} and consequences of
the choice of language and decomposition in that language are further
discussed in more detail in Section~\ref{sec:discussion}.

\paragraph{Distributed proxies}
Our use of program decomposition provides for partial handling of
\emph{distributed representations}, the idea that concepts can be
distributed among multiple entities.
In our case, influence and association of a protected information type can be
distributed among multiple program points.
First, substitution (denoted by $[p_1/X]p_2$) is defined to
replace \emph{all} instances of variable $X$ in $p_2$ with the program
$p_1$.
If there are multiple instances of $ X $ in $ p_2 $, they are still
describing a single decomposition and thus the multiple instances of
$ p_2 $ in $p_1$ are viewed as a single proxy.
Further, implementations of substitution can be (and is in our
implementation) associativity-aware: programs like $x_1 + x_2 + x_3$
can be equivalent regardless of the order of the expressions in that they
can be decomposed in exactly the same set of ways.
If a proxy is distributed among $x_1$ and $x_3$, it will still be
considered by our methods because $ x_1 + (x_2 + x_3) $ is equivalent
to $ (x_1 + x_3) + x_2 $, and the sub-expression $ x_1 + x_3 $ is part
of a valid decomposition. Allowing such equivalences within the implementation
of substitution partially addresses the problem that our theory does
not respect semantic equivalence, which is a necessary consequence
of Theorem~\ref{thm:sem-impossibility}.

\begin{figure}
\begin{grammar}
<aexp> ::= $\mathbb{R}$ | var | op(<aexp>$, \ldots,$ <aexp>)
  \alt $\mathbf{ite}$(<bexp>, <aexp>, <aexp>)

<bexp> ::= T | F | $\lnot$ <bexp>
  \alt op(<bexp>$, \ldots,$ <bexp>)
  \alt relop(<aexp>, <aexp>)

<prog> ::= $\lambda \mathrm{var}_1, \ldots, \mathrm{var}_n$ . <aexp>
\end{grammar}

\caption{Syntax for the language used in our analysis.}
\label{fig:language}
\end{figure}

\subsection{Analyzing Proxy Use} \label{sec:detecting-analyzing}
Algorithm~\ref{alg:detect-generic-informal} describes a general
technique for detecting \edproxyuse in expression programs.
In addition to the parameters and expression, it takes as input a
description of the distribution governing the feature variables $X$
and $Z$.
In practice this will nearly always consist of an empirical sample,
but for the sake of presentation we simplify here by assuming the
distribution is explicitly given.
In Section~\ref{sec:estimates}, we describe how the algorithm can
produce estimates from empirical samples.

The algorithm proceeds by enumerating sub-expressions of the given
program.
For each sub-expression $e$ appearing in $p$, $\Call{ProxyDetect}{}$
computes the set of positions at which $e$ appears.
If $e$ occurs multiple times, we consider all possible subsets of
occurrences as potential
decompositions\ifdefined\shortappendix\else\footnote{This occurs
  often in decision forests (see Figure~\ref{fig:big}).} \fi.
It then iterates over all combinations of these positions, and creates
a decomposition for each one to test for \edproxyuse.
Whenever the provided thresholds are exceeded, the decomposition is
added to the return set.
This proceeds until there are no more subterms to consider.
While not efficient in the worst-case, this approach is both sound and
complete with respect to Definition~\ref{def:quant-usage}.

\begin{thm}[Detection soundness]
  \label{thm:detect-sound} Any decomposition $(p_1, p_2) $ returned by
  $ \Call{ProxyDetect}{p, \features, \epsilon, \delta}$ is a
  decomposition of the input program $ p $ and had to pass the
  $ \epsilon,\delta $ thresholds, hence is a \edproxyuse.
\end{thm}

\begin{thm}[Detection completeness]\label{thm:detect-complete}
  Every decomposition which could be a \edproxyuse is enumerated by
  the algorithm.
  Thus, if $(p_1, p_2)$ is a decomposition of $p$ with
  $\iota(p_1, p_2) \geq d$ and
  $d(\llbracket p_1 \rrbracket(\features), Z) \geq \epsilon$, it will
  be returned by $\Call{ProxyDetect}{p, \features, \epsilon, \delta}$.
\end{thm}

Our detection algorithm considers single terms in its decomposition.
Sometimes a large number of syntactically different proxies with weak
influence might collectively have high influence.
A stronger notion of program decomposition that allows a collection of
multiple terms to be considered a proxy would identify such a case of
proxy use but will have to search over a larger space of expressions.
Exploring this tradeoff between scalability and richer proxies is an
important topic for future work.

The detection algorithm runs in time
$ \bigO{\setsize{\prog} c \paren{\setsize{\dataset} + k
    \setsize{\dataset}}} $ where $ \setsize{\dataset} $ is the size of
a dataset employed in the analysis, $ c $ is the number of
decompositions of a program, $ k $ is the maximum number of elements
in the ranges of all sub-programs ($\setsize{\dataset}$ in the worst
case), and $ \setsize{\prog} $ is the number of sub-expressions of a
program.
The number of decompositions varies from $\bigO{\setsize{\prog}} $ to
$ \bigO{2^{\setsize{\prog}}} $ depending on the type of program
analyzed.
Details can be found in Appendix~\ref{appendix:complexity} along with
more refined bounds for several special cases.

\section{Removing Proxy Use Violations} \label{sect:repair}

\newcommand{\lref}[1]{Line~\ref{#1}}

In this section we present a repair algorithm for removing violations of
\edproxyusage in a model.  Our approach has two parts: first
(Algorithm~\ref{alg:repair-informal}) is the iterative discovery of proxy uses
via the $\Call{ProxyDetect}{}$ procedure described in the previous section and
second (Algorithm~\ref{alg:repair-local-informal}) is the repair of the ones
found by the oracle to be violations. We describe these algorithms informally
here, and Appendix~\ref{app:repair} contains formal descriptions of these
algorithms. The iterative discovery procedure guarantees that the returned
program is free of violations (Algorithm~\ref{alg:repair}).  Our repair
procedures operate on the expression language, so they can be applied to any
model that can be written in the language.  Further, our violation repair
algorithm does not require knowledge of the training algorithm that produced
the model.  The  witnesses of proxy use localize where in the program
violations occur. To repair a violation we search through expressions
\emph{local} to the violation, replacing the one which has the least impact on
the accuracy of the model that at the same time reduces the association or
influence of the violation to below the $ (\epsilon,\delta) $ threshold.

At the core of our violation repair algorithm is the simplification of
sub-expressions in a model that are found to be violations.
Simplification here means the replacement of an expression that is not
a constant with one that is.
Simplification has an impact on the model's performance hence we take
into account the goal of preserving utility of the machine learning
program we repair.
We parameterize the procedure with a measure of utility $v$ that
informs the selection of expressions and constants for simplification.
We briefly discuss options and implementations for this parameter
later in this section.

The repair procedure (Algorithm~\ref{alg:repair-local-informal}) works as
follows.
Given a program $ p $ and a decomposition $(p_1,p_2) $, it first finds
the best simplification to apply to $ p $ that would make
$ (p_1,p_2) $ no longer a violation.
This is done by enumerating expressions that are \emph{local} to
$ p_1 $ in $ p_2 $ (\lref{line:repair-local}).
Local expressions are sub-expressions of $ p_1 $ as well as $ p_1 $
itself and if $ p_1 $ is a guard in an if-then-else expression, then
local expressions of $ p_1 $ also include that if-then-else's true and
false branches as well as their sub-expressions.
Each of the local expressions corresponds to a decomposition of $ p $
into the local expression $ p_1' $ and the context around it $ p_2' $.
For each of these local decompositions we discover the best constant,
in terms of utility, to replace $ p_1' $ with
(\lref{line:repair-optimal}).
We then make the same simplification to the original decomposition
$(p_1, p_2)$, resulting in $ (p_1'',p_2'') $
(\lref{line:repair-subst}) Using this third decomposition we check
whether making the simplification would repair the original violation
(\lref{line:repair-violation}), collecting those simplified programs
that do.
Finally, we take the best simplification of those found to remove the
violation (\lref{line:repair-optimal2}). Details on how the
optimal constant is selected is described in Appendix~\ref{app:optimal-constant}.

\begin{algorithm}[t]
\caption{Witness-driven repair.}
\begin{algorithmic}
\Require association (\assocmeasure{}), influence (\infmeasure{}), utility (\utilmeasure{}) measures, oracle ($\oracle{}$)
\Procedure{Repair}{$p, \features, Z, \epsilon, \delta$}
\State $ P  \gets \set{d \in \Call{ProxyDetect}{p,\features,Z,\epsilon, \delta} : \text{ not } \oracle(d)} $
\If{ $ P \neq \emptyset $ }
\State $(p_1,p_2) \gets \text{element of } P $
\State $p' \gets \Call{ProxyRepair}{p, (p_1,p_2), \features,Z,\epsilon, \delta} $
\State \Return $\Call{Repair}{p',\features, Z, \epsilon, \delta} $
\Else
\State \Return $p$
\EndIf
\EndProcedure
\end{algorithmic}
\label{alg:repair-informal}
\end{algorithm}

\begin{algorithm}[t]
\caption{Local Repair.}
\begin{algorithmic}[1]
\Require association (\assocmeasure{}), influence (\infmeasure{}), utility (\utilmeasure{}) measures
\Procedure{ProxyRepair}{$p, (p_1,p_2), \features, Z, \epsilon, \delta$}
\State $R \gets \set{}$
\For{each subprogram $p_1'$ of $p_1$}\label{line:repair-local}
\State $r^* \gets$~Optimal constant for replacing $p_1'$ \label{line:repair-optimal}
\State $(p_1'',p_2'') \gets (p_1,p_2) $ with $ r^* $ subst. for $ p_1' $ \label{line:repair-subst}
\If{$\iota(p_1'', p_2'') \leq \delta \lor d(\llbracket p_1'' \rrbracket(\features), Z) \leq \epsilon$} \label{line:repair-violation}
\State $ R \gets R \cup \subst{u}{r^*}{p_2'}$
\EndIf
\EndFor
\State \Return $\argmax_{p^*\in R} \utilmeasure\paren{p^*}$ \label{line:repair-optimal2}
\EndProcedure
\end{algorithmic}
\label{alg:repair-local-informal}
\end{algorithm}

Two important things to note about the repair procedure.
First, there is always at least one subprogram on
\lref{line:repair-local} that will fix the violation, namely the
decomposition $(p_1,p_2)$ itself.
Replacing $ p_1 $ with a constant in this case would disassociate it
from the sensitive information type.
Secondly, the procedure produces a model that is smaller than the one
given to it as it replaces a non-constant expression with a constant.
These two let us state the following:

\begin{thm}\label{thm:repair}
  Algorithm~\ref{alg:repair-informal} terminates and returns a program
  that does not have any \edproxyusage violations (instances of
  \edproxyusage for which oracle returns false).
\end{thm}

\section{Evaluation}
\label{sect:eval}

In this section we empirically evaluate our definition and algorithms
on several real datasets. 
In particular, we simulate a financial services application and
demonstrate a typical workflow for a practitioner using our tools to
detect and repair proxy use in decision trees and linear models
(\S\ref{sect:eval-workflow}). 
We highlight that this workflow identifies more proxy uses over a
baseline procedure that simply removes features associated with a
protected information type. 
For three other simulated settings on real data sets---contraception
advertising, student assistance, and credit advertising---we describe
our findings of interesting proxy uses and demonstrate how the outputs
of our detection tool would allow a normative judgment oracle to
determine the appropriateness of proxy uses (\S\ref{sect:eval-cases}). 
\ifdefined\shortappendix In \S\ref{sect:eval-detection}, \else In
\S\ref{sect:eval-detection}, we evaluate the performance of our
detection and repair algorithms and show that in particular cases, the
runtime of our system scales linearly in the size of the model. 
Also, \fi by injecting violations into real data sets so that we have
ground truth, we evaluate the completeness of our algorithm, and show
a graceful degradation in accuracy as the influence of the violating
proxy increases.

\paragraph{Models and Implementation}
Our implementation currently supports linear models, decision trees,
random forests, and rule lists.
Note that these model types correspond to a range of commonly-used
learning algorithms such as logistic regression, support vector
machines~\cite{cortes95svm}, CART~\cite{Breiman01}, and Bayesian rule
lists~\cite{letham2015}.
Also, these models represent a significant fraction of models used in
practice in predictive systems that operate on personal information,
ranging from advertising~\cite{chickering00advertising},
psychopathy~\cite{hare03pclr}, criminal
justice~\cite{berk14forecasts,berk16forecasting}, and actuarial
sciences~\cite{frees2014asbook,gepp16insurance}.
Our prototype implementation was written in Python, and we use
scikit-learn package to train the models used in the evaluation.
\ifdefined\evenshorterappendix\else The benchmarks we describe later in
this section were recorded on a Ubuntu Desktop with 4.2 GHz Intel Core
i7 and 32GB RAM. \fi

\subsection{Example Workflow}
\label{sect:eval-workflow}

A financial services company would like to expand its client base by
identifying potential customers with high income. To do so, the company hires
an analyst to build a predictive model that uses age, occupation, education
level, and other socio-economic features to predict whether an individual
currently has a ``high'' or ``low'' income. This practice is in line with the
use of analytics in the financial industry that exploit the fact that
high-income individuals are more likely to purchase financial
products~\cite{soft-finance}.

Because demographic data is known to correlate with marital
status~\cite{marriage-income}, the data processor would like to ensure that
the trained model used to make income predictions does not effectively infer
individuals' marital status from the other demographic variables that are
explicitly used. In this context, basing the decision of which clients to
pursue on marital status could be perceived as a privacy violation, as other
socio-economic variables are more directly related to one's interest and
eligibility in various financial services.

To evaluate this scenario, we trained an income prediction model from
the UCI Adult dataset which consists of roughly 48,000 rows containing
economic and demographic information for adults derived from
publicly-available U.S.
Census data.
One of the features available in this data is marital status, so we
omitted it during training, and later used it when evaluating our
algorithms. 
In this scenario, we act as the oracle in order to illustrate the kind
of normative judgments an analyst would need to make as an oracle.

\begin{figure}[t]
  \centering\includegraphics[width=\linewidth]{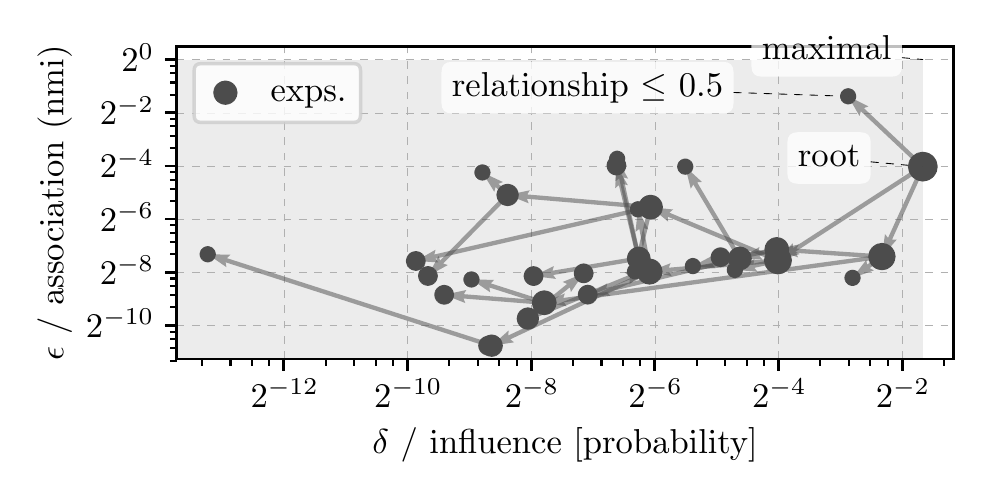}
\caption{\label{fig:plot_adult_tree_subterms} The association and
  influence of the expressions composing a decision
  tree trained on the UCI Adult dataset.
  Narrow lines designate the sub-expression relationship. Shaded area
  designates the feasible values for association and influence between
  none, and maximal. Marker size denotes the relative size of the
  sub-expressions pictured.
}
\end{figure}

After training a classifier on the preprocessed dataset, we found a
strong proxy for marital status in terms of an expression involving
relationship status. 
Figure~\ref{fig:plot_adult_tree_subterms} visualizes all of the
expressions making up the model (marked as $\bullet$), along with
their association and influence measures.
In decision trees, sub-expressions like these coincide with
decompositions in our proxy use definition; each sub-expression can be
associated with a decomposition that cuts out that sub-expression from
the tree, and leaves a variable in its place. 
The connecting lines in the figure denote the sub-expression
relationship. 
Together with the placement of points on the influence and association
scales, this produces an overview of the decision tree and the
relationship of its constituent parts to the sensitive attribute.

\begin{figure}[t]
\centering\includegraphics[width=\linewidth]{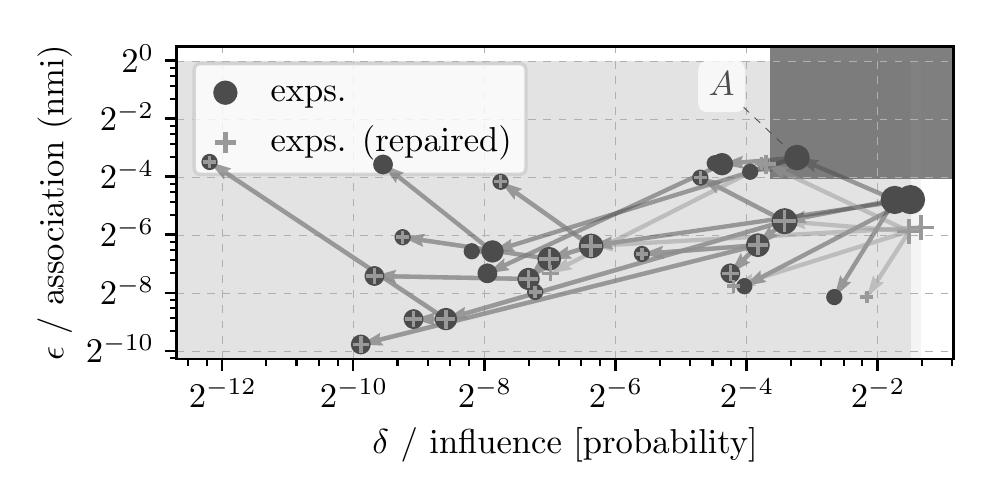}
\caption{\label{fig:plot_adult_tree_post_removal}
  Decision tree trained on the UCI Adult dataset but with the
  $\texttt{relationship}$ attribute removed ($\bullet$), and the repaired
  version ($+$) of the same tree. Dark area in the upper-left designates
  the thresholds used in repair.
}
\end{figure}

On further examination the relationship status was essentially a
finer-grained version of marital status.
While not interesting in itself, this occurrence demonstrates an issue
with black-box use of machine learning without closely examining the
structure of the data.
In particular, one can choose to remove this feature, and the model
obtained after retraining will make predictions that have low
association with marital status.
However, one submodel demonstrated relatively strong \proxyuse
$(\epsilon = 0.1, \delta = 0.1)$:
$\mathtt{age} \leq 31 ~\text{and}~\mathtt{sex} = 0 ~\text{and}~
\mathtt{capital\_loss} \leq 1882.50$\ (labeled A in
Figure~\ref{fig:plot_adult_tree_post_removal}).
This demonstrates that simply removing a feature does not ensure that
proxies are removed.
When the model is retrained, the learning algorithm might select new
computations over other features to embed in the model, as it did in
this example.
Also, note that the new proxy combines three additional features.
Eliminating all of these features from the data could impact model
performance.
Instead we can use our repair algorithm to remove the proxy: we
designate the unacceptable $\epsilon,\delta$ thresholds (the darkest
area in Figure~\ref{fig:plot_adult_tree_post_removal}) and repair any
proxies in that range.
The result is the decision tree marked with $+$ in the figure.
Note that this repaired version has no sub-expressions in the
prohibited range and that most of the tree remains unchanged (the
$\bullet$ and $+$ markers largely coincide).

\subsection{Other Case Studies}
\label{sect:eval-cases}

We now briefly discuss interesting examples for proxy use from other
case studies, demonstrating how our framework aids normative use
privacy judgments. 
\ifdefined\shortappendix\else More details on these datasets and
experiments are in Appendix~\ref{app:case-study-details}. \fi

\paragraph{Targeted contraception advertising} We consider a scenario
in which a data processor wishes to show targeted advertisements for
contraceptives to females.
We evaluated this scenario using data collected for the 1987 National
Indonesia Contraceptive Survey~\cite{indonesia-contra}, which contains
a number of socio-economic features, including feature indicating
whether the individual's religious beliefs were Islam.
A decision tree trained on this dataset illustrates an interesting
case of potential use privacy via the following proxy for religion:
$ ite(\mathtt{educ} < 4 \land \mathtt{nchild} \leq 3 \land
\mathtt{age} < 31, \mathtt{no}, \mathtt{yes}) $.
This term predicts that women younger than 31, with below-average
education background and fewer than four children will not use
contraception. 
In fact, just the ``guard'' term $\mathtt{educ} < 4$ alone is more
closely associated with religion, and its influence on the model's
output is nearly as high. 
This reveals a surprising association between education levels and
religion leading to a potentially concerning case of proxy use.

\paragraph{Student assistance} A current trend in education is the use of predictive analytics to identify students who are likely to benefit from certain types of interventions~\cite{uc-analytics, k12-analytics}. We look at a scenario where a data processor builds a model to predict whether a secondary school student's grades are likely to suffer, based on a range of demographic features, social information, and academic information.
To evaluate this scenario, we trained a model on the UCI Student Alcohol
Consumption dataset~\cite{student-alcohol}, with alcohol use as the sensitive
feature.  Our algorithm found the following proxy for alcohol use:
$\mathtt{studytime} < 2$. This finding suggests that this instance of proxy use
can be deemed an appropriate use, and not a privacy violation, as the amount of
time a student spends studying is clearly relevant to their academic
performance.

\paragraph{Credit advertisements} We consider a situation where a credit card
company wishes to send targeted advertisements for credit cards based on
demographic information. In this context, the use of health status for targeted
advertising is a legitimate privacy concern~\cite{dixon14scores}.
To evaluate this scenario, we trained a model to predict interest in
credit cards using the PSID dataset. From this, we trained two models: one that identifies individuals with
student loans and another that identifies individuals with existing
credit cards as the two groups to be targeted.
The first model had a number of instances of proxy use.
One particular subcomputation that was concerning was a subtree of the
original decision tree that branched on the number of children in the
family.
This instance provided negative outcomes to individuals with more
children, and may be deemed inappropriate for use in this context.
In the second model, one proxy was a condition involving income
$\mathtt{income} \leq 33315$.
The use of income in this context is justifiable, and therefore this
may be regarded as not being a use privacy violation.

\subsection{Detection and Repair}\label{sect:eval-detection}

\ifdefined\evenshorterappendix\else

\begin{figure}[t]
\centering\includegraphics[width=\linewidth]{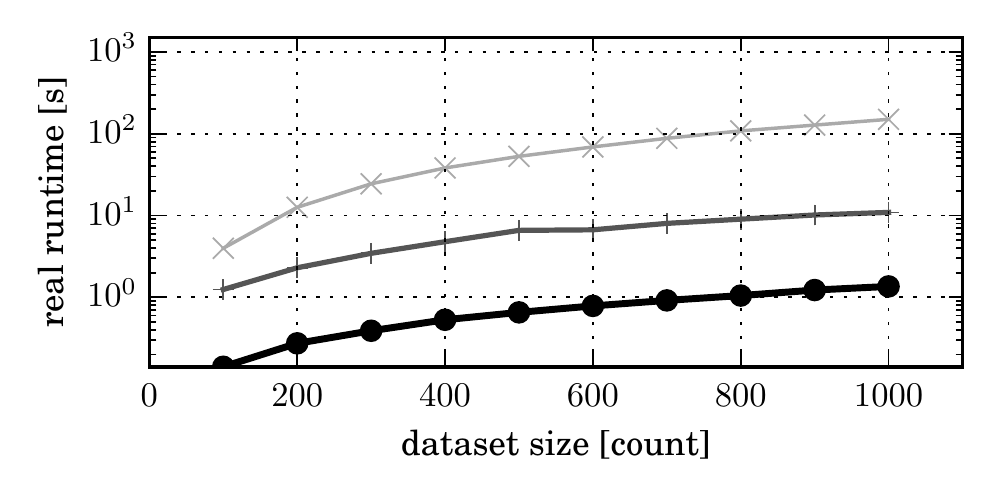}
\caption{\label{fig:bench-detect} Worst-case detection algorithm
  run-time (average of 5 runs) as a function of input dataset size.
  Influence and association computed on each decomposition (hence
  worst-case).
  The models are decision tree($\circ$), random forest($+$), and
  logistic regression($\times$) trained on the UCI Adult dataset.}
\end{figure}

For the remainder of the section we focus on evaluating the performance
and efficacy of the detection and repair algorithms.
We begin by exploring the impact of the dataset and model
size on the detection algorithm's runtime.

Figure~\ref{fig:bench-detect} demonstrates the runtime of our
detection algorithm on three models trained on the UCI Adult dataset
vs.
the size of the dataset used for the association and influence
computations.
The algorithm here was forced to compute the association and influence
metrics for each decomposition (normally influence can be skipped if
association is below threshold) and thus represents a worst-case
runtime.
The runtime for the random forest and decision tree scales linearly in
dataset size due to several optimizations.
The logistic regression does not benefit from these and scales
quadratically.
Further, runtime for each model scales linearly in the number of
decompositions \ifdefined\shortappendix\else(see
Appendix~\ref{sec:bench-detect-model-size}) \fi, but logistic
regression models contain an exponential number of decompositions as a
function of their size.

\fi

\newcommand{\attrib}[1]{$\mathtt{#1}$}

\begin{figure}[t]
\centering\includegraphics[width=\linewidth]{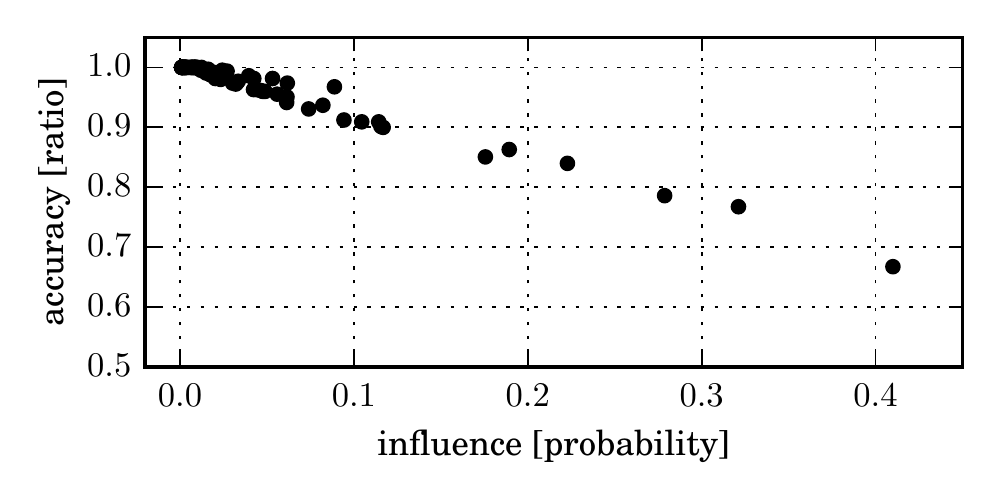}
\caption{\label{fig:plot_repair_random} Repaired accuracy vs.
  influence of proxy during repair of a synthetic proxy inserted into
  random positions of a decision tree trained on the UCI Student
  Alcohol Consumption dataset.
  Accuracy is agreement to non-repaired model.
  The synthetic model is a $(1.0)$-proxy for alcohol use, inserted
  into a decision tree predicting student grade.
  Repair is configured for \argproxyuse{0.01}{0.01} removal.
  Note that other proxies (if they exist) are not repaired in this
  experiment.}
\end{figure}

To determine the completeness of our detection algorithm we inserted a
proxy in a trained model to determine whether we can detect it.
To do this, we used the UCI Student Alcohol Consumption dataset to
train two decision trees: one to predict students' grades, and one to
predict alcohol consumption.
We then inserted the second tree into random positions of the first
tree thereby introducing a proxy for alcohol consumption.
We observed that in each case, we were able to detect the introduced
proxy.
While not interesting in itself due to our completeness theorem, we
used this experiment to explore how much utility is actually lost due
to repair.
We evaluate our repair algorithm on a set of similar models with
inserted violations of various influence magnitude.
The results can be seen in Figure~\ref{fig:plot_repair_random}.
We can see that the accuracy (i.e., ratio of instances that have
agreement between repaired and unrepaired models) falls linearly with
the influence of the inserted proxy.
This implies that repair of less influential proxies will incur a
smaller accuracy penalty than repair of more influential proxies.
In other words, our repair methods do not unduly sacrifice accuracy
when repairing only minor violations.

A point not well visible in this figure is that occasionally repair
incurs no loss of utility.
This is due to our use of the scikit-learn library for training
decision trees as it does not currently support pruning unnecessary
nodes.
Occasionally such nodes introduce associations without improving the
model's accuracy.
These nodes can be replaced by constants without loss.
We have also observed this in some of our case studies.

\section{Related Work}
\label{sect:related}

\subsection{Definition}

\paragraph{Minimizing disclosures}
In the computer science literature, privacy has been thought of as the
ability to protect against undesired flows of information to an
adversary.
Much of the machinery developed in cryptography, such as encryption,
anonymous communication, private computation, and database privacy
have been motivated by such a goal.
Differential privacy~\cite{dwork06crypto} is one of the main pillars
of privacy research in the case of computations over data aggregated
from a number of individuals, where any information gained by an
adversary observing the computation is not caused by an individual's
participation.
However, none of these technologies cover the important setting of
individual-level data analytics, where one may want to share some
information while hiding others from adversaries with arbitrary
background knowledge.
This absence is with good reason, as in the general case it is
impossible to prevent flows of knowledge from individual-level data,
while preserving the utility of such data, in the presence of
arbitrary inferences that may leverage the background knowledge of an
adversary~\cite{dwork06icalp}.
In this work, we do not attempt to solve this problem either.

Nevertheless, the setting of individual level data analytics is
pervasive, especially in the case of predictive systems that use
machine learning.
Since these systems are largely opaque, even developers do not have
a handle on information they may be inadvertently using via
inferences.
Therefore, in this work, we make the case for proxy use restrictions
in data driven systems and develop techniques to detect and repair
violations of proxy use.
Restrictions on information use, however do not supplant the need for
other privacy enhancing technologies geared for restricting information
collection and disclosure, which may be useful in conjunction with the
enforcement of use restrictions.
For example, when machine learning models are trained using personal
data, it is desirable to minimize disclosures pertaining to
individuals in the training set, and to reduce the use of protected
information types for the individuals the models are applied to.

\paragraph{Identifying explicit use}
The privacy literature on use restrictions has typically focused on explicit
use of protected information types, not on proxy use
(see Tschantz et al.~\cite{tschantz12sp}
for a survey and Lipton and Regan~\cite{lipton16privacy}). Recent work on
discovering personal data use by black-box web services focuses mostly on
explicit use of protected information types by examining causal effects
\cite{datta15pets,sunlight}; some of this work also examines
associational effects~\cite{xray,sunlight}. Associational effects
capture some forms of proxy use but not others as we argued in
Section~\ref{sec:defn}.

\subsection{Detection and Repair Models}

Our detection algorithm operates with white-box access to the prediction model.
Prior work requires weaker access assumptions.

\paragraph{Access to observational data} Detection techniques working
under an associative use definition~\cite{fairtest,
  feldman15disparate} usually only require access to observational
data about the behavior of the system.

\paragraph{Access to black-box experimental data} Detection techniques
working under an explicit use definition of information
use~\cite{sunlight,datta15pets} typically require experimental
access to the system. This access allows the analyst to control some inputs to
the system and observe relevant outcomes.

The stronger white-box access level allows us to decompose the model and trace an
intermediate computation that is a proxy.
Such traceability is not afforded by the weaker access assumptions in
prior work. Thus, we explore a different point in the space by
giving up on the weaker access requirement to gain the ability to
trace and repair proxy use.

Tram\`{e}r et al.~\cite{fairtest} solve an important orthogonal problem of
efficiently identifying populations where associations may appear.  Since our
definition is parametric in the choice of the population, their technique
could allow identifying relevant populations for further analysis
using our methods.

\paragraph{Repair}
Removal of violations of privacy can occur at different
points of the typical machine learning pipeline.
Adjusting the training dataset is the most popular approach, including variations that
relabel only the class attribute~\cite{luong11knn}, modify entire instances
while maintaining the original schema~\cite{feldman15disparate}, and transform
the dataset into another space of features~\cite{zemel13fair,dwork11fairness}.
Modifications to the training algorithm are specific to the trainer employed (or to a class
of trainers). Adjustments to Naive Bayes~\cite{calders10naive} and trainers
amiable to regularization~\cite{kamishima11fairness} are examples. Several techniques for producing
differentially-private machine learning models modify trained models by perturbing coefficients~\cite{CMS11,BST14}.
Other differentially-private data analysis
techniques~\cite{DMNS06} instead perturb the output by adding
symmetric noise to the true results of statistical queries.
All these repair techniques aim to minimize associations or inference
from the outcomes rather than constrain use.

\section{Discussion}
\label{sec:discussion}

\paragraph{Beyond strict decomposition}
Theorem~\ref{thm:sem-impossibility} shows that a definition
satisfying natural semantic properties is impossible.
This result motivates our syntactic definition, parameterized by a
programming language and a choice of program decomposition.
In our implementation, the choice of program decomposition is strict. It only
considers single terms in its decomposition. However, proxies may be
distributed across different terms in the program.
As discussed in Section~\ref{sec:detecting:model}, single term decompositions
can also deal with a restricted class of such distributed proxies.
Our implementation does not identify situations
where each of a large number of syntactically different proxies have weak
influence but together combine to result in high influence.
A stronger notion of program
decomposition that allows a collection of multiple terms to be
considered a proxy would identify such a case of proxy use.

The choice of program decomposition also has consequences for the
tractability of the detection and repair algorithms.
The detection and repair algorithms presented in this paper currently
enumerate through all possible subprograms in the worst case.
Depending on the flexibility of the language chosen and the
model\footnote{Though deep learning models can be expressed in the
  example language presented in this paper, doing so would result in
  prohibitively large programs.}
being expressed there could be an exponentially large number of
subprograms, and our enumeration would be intractable.

Important directions of future work are therefore organized along two
thrusts.
The first thrust is to develop more flexible notions of program
decompositions that identify a wide class of proxy uses for other
kinds of machine learning models, including deep learning models that
will likely require new kinds of abstraction techniques due to their
large size.
The second thrust is to identify scalable algorithms for detecting and
repairing proxy use for these flexible notions of program
decompositions.

\paragraph{Data and access requirements}
Our definitions and algorithms require (i) a specification of which
attributes are protected, (ii) entail reasoning using data about these protected information types for individuals, and
(iii) white box access to models and a representative dataset of inputs.
Obtaining a complete specification of protected information types can be challenging
when legal requirements and privacy expectations are vague regarding protected
information types. However, in many cases, protected types are
specified in laws and regulations governing the system under study (e.g.,
HIPAA, GDPR), and also stated in the data processor’s privacy policies.

Further, data about protected information types is often not explicitly collected.
Pregnancy status, for example, would rarely find itself as an explicit
feature in a purchases database (though it was the case in the Target
case).
Therefore, to discover unwanted proxy uses of protected information types, an
auditor might need to first infer the protected attribute from the
collected data to the best extent available to them.
Though it may seem ethically ambiguous to perform a protected
inference in order to (discover and) prevent protected inferences, it
is consistent with the view that privacy is a function of both
information and the purpose for which that information is being used
\cite{TschantzDW12}\footnote{This principle is exemplified by law in
  various jurisdictions including the PIPEDA Act in Canada
  \cite{pipeda}, and the HIPAA Privacy Rule in the USA \cite{hipaa}.}.
In our case, the inference and use of protected information by an
auditor has a different (and ethically justified) purpose than
potential inferences in model being audited.
Further, protected information has already been used by public and
private entities in pursuit of social good: affirmative action
requires the inference or explicit recording of minority membership,
search engines need to infer suicide tendency in order to show suicide
prevention information in their search results\cite{suicide}, health
conditions can potentially be detected early from search logs of
affected individuals \cite{paparrizos16screening}.
Supported by law and perception of public good, we think it justified
to expect system owners be cooperative in providing the necessary
information or aiding in the necessary inference for auditing.

Finally, in order to mitigate concerns over intellectual property due to access
requirements for data and models, the analyst will need to be an internal
auditor or trusted third party; existing privacy-compliance audits (Sen et al.
~\cite{sen14sp}) that operate under similar requirements could be augmented with our methods.

\paragraph{Normative judgments}

Appropriateness decisions by the analyst will be made in accordance with legal
requirements and ethical norms. Operationally, this task might fall on privacy
compliance teams. In large companies, such teams include law, ethics, and
technology experts. Our work exposes the specific points where these complex
decisions need to be made.
In our evaluation, we observed largely
human-interpretable witnesses for proxies. For more complex models, additional
methods from interpretable machine learning might be necessary to make
witnesses understandable.

Another normative judgment is the choice of acceptable $\epsilon, \delta$
parameters.  Similar to differential privacy, the choice of parameters requires
identifying an appropriate balance between utility and privacy.  Our
quantitative theory could provide guidance to the oracle on how to prioritize
efforts, e.g., by focusing on potentially blatant violations (high $\epsilon,
\delta$ values).

\section{Conclusion}
\label{sect:conclusion}

We develop a theory of use privacy in data-driven systems.
Distinctively, our approach constrains not only the
direct use of protected information types but also their proxies (i.e.
strong predictors), unless allowed by exceptions justified by ethical
considerations.

We formalize proxy use and present a program analysis technique for
detecting it in a model.
In contrast to prior work, our analysis is white-box.
The additional level of access enables our detection algorithm to
provide a witness that localizes the use to a part of the algorithm.
Recognizing that not all instances of proxy use of a protected
information type are inappropriate, our theory of use privacy makes
use of a normative judgment oracle that makes this appropriateness
determination for a given witness.
If the proxy use is deemed inappropriate, our repair algorithm uses
the witness to transform the model into one that does not exhibit
proxy use.
Using a corpus of social
datasets, our evaluation shows that these algorithms are able to detect proxy
use instances that would be difficult to find using existing techniques, and
subsequently remove them while maintaining acceptable classification
performance.

\paragraph*{Acknowledgments}
We would like to thank Amit Datta, Sophia Kovaleva, and Michael C.
Tschantz for their thoughtful discussions throughout the development
of this work. We thank our shepherd Aylin Caliskan and
anonymous reviewers for their numerous suggestions that improved this paper.

\begin{small}
  This work was developed with the support of NSF grants CNS-1704845,
  CNS-1064688 as well as by DARPA and the Air Force Research
  Laboratory under agreement number FA8750-15-2-0277.
  The U.S.
  Government is authorized to reproduce and distribute reprints for
  Governmental purposes not withstanding any copyright notation
  thereon.
  The views, opinions, and/or findings expressed are those of the
  author(s) and should not be interpreted as representing the official
  views or policies of DARPA, the Air Force Research Laboratory, the
  National Science Foundation, or the U.S.
  Government.
\end{small}

\bibliographystyle{styles/ACM-Reference-Format}
\bibliography{%
  bibs/common.bib,%
  bibs/security.bib,%
  bibs/pl.bib,%
  bibs/other.bib,%
  bibs/causal-influence.bib,%
  bibs/blackbox.etc.bib%
}

\appendix

\algtext{EndIf}{{\bf end if}}
\algtext{EndFor}{{\bf end for}}

\section{Proof of Theorem~$\ref{thm:sem-impossibility}$}
\label{app:impossibility-proof}

\noindent
{\bf Theorem 1.} {\it
No definition of proxy use can satisfy Properties \ref{axm:explicit-sem}-\ref{axm:indep-sem} simultaneously.}

\begin{proof}
Proof by contradiction. Assume that a definition of proxy use
satisfies all four properties. Let $X$, $Y$, and $Z$ be uniform binary random
variables, such that $\Pr(Y = X \oplus Z) = 1$, but $X$, $Y$ and $Z$ are pairwise independent.
By (explicit use of proxy), the model $\model(Y, Z) = Y\oplus Z$ has proxy use of $Z$.
By (dummy), the model $\model'(Y, Z, X) = Y\oplus Z$ has proxy use of $Z$. Choose $f(x, z) = x\oplus z$.
By our assumption earlier, $\prob{Y = f(X, Z)} = 1$.
Therefore, by (preprocessing), the model $\model''(Z, X) = \model'(f(X, Z), Z, X)$ has proxy use of $Z$.
Note that $\model''(Z, X) = X\oplus Z\oplus Z = X$. Therefore, by (dummy), $\model'''(X) = X$ has proxy use of $Z$. But,
by (independence), $\model'''$ does not have proxy use of $Z$. Therefore, we have a contradiction.
\end{proof}

The key intuition behind this result is that Property \ref{axm:prep-sem} requires proxy
use to be preserved when an input is replaced with a function that
predicts that input via composition. However, with a purely semantic view of function
composition, the causal effect of the proxy can disappear. The particular
example of this observation we use in the proof is $Y\oplus Z$, where $Z$ is the protected
information type.
This function has proxy use of $Z$. However, if $X\oplus Z$ is a
perfect predictor for $Y$, then the example can be reduced to $X\oplus Z\oplus
Z = X$, which has no proxy use of $Z$. To overcome this impossibility
result, we choose a more syntactic notion of function composition, which is
tied to how the function is represented as a program, and looks for evidence
of proxy use within the representation.

\section{Algorithm for Detection}
\label{app:detection}

In this section we provide technical details about the detection
algorithm skipped from the main body of the paper.
In particular, we formally define the decomposition used in the
implementation, how machine learning models are translated to the term
language, and how associational tests mitigate spurious results due to
sampling.

\subsection{Decomposition}
Before we present the formal algorithm for detection, we need to
develop notation for precisely denoting decompositions.
Decomposition follows naturally from the subterm relation on
expressions.
However, as identical subterms can occur multiple times in an
expression, care must be taken during substitution to distinguish
between occurrences.
For this reason we define substitution positionally, where the subterm
of expression $e = \mathrm{op}(e_1, \ldots, e_n)$ at position $q$,
written $e |_q$, is defined inductively:
\[
\mathrm{op}(e_1, \ldots, e_n) |_q
=
\left\{
\begin{array}{ll}
\mathrm{op}(e_1, \ldots, e_n) & \mathrm{if\ } q = \epsilon \\
e_i |_{q'} & \mathrm{if\ } q = iq' \land 1 \le i \le n \\
\mathrm{op}(e_{i_1},\ldots, e_{i_k}) & \mathrm{if\ } q = \{i_1,\ldots,i_k\} \\
\perp & \mathrm{otherwise}
\end{array}
\right.
\]
We denote $q$ as `positional indicator'.
Specifically, $q$ has the syntax of the following.
\begin{grammar}
<q> ::= $\epsilon$ | $i$<q> | $\{i_1,\ldots,i_k\}$
\end{grammar}
We then define the term obtained by substituting $s$ in $e$ at
position $q$, written $e[s]_q$, to be the term where $e[s]_q |_q = s$,
and $e[s]_q |_{q'} = e_{q'}$ for all $q'$ that are not prefixed by
$q$.
For a sequence of positions $q_1, \ldots, q_n$ and terms
$s_1, \ldots, s_n$, we write $e[s_1, \ldots, s_n]_{q_1, \ldots, q_n}$
to denote the sequential replacement obtained in order from $1$ to
$n$.
Given a program $p = \lambda \vec{x} .
e$, we will often write $p |_q$ or $p[s]_q$ for brevity to refer to
$e |_q$ and $e[s]_q$, respectively.
The set of decompositions of a program $p$ is then defined by the set
of positions $q$ such that $p |_q \ne \perp$.
Given position $q$, the corresponding decomposition is simply
$(\lambda \vec{x} .
p |_q, u, \lambda \vec{x}, u .
p[u]_q)$.

\begin{example}
Consider a simple model,
\begin{align*}
p &= \lambda x, y . \mathbf{ite}(x+y \le 0, 1, 0) \\
  &= \lambda x, y . \mathbf{ite}(\le(+(x, y), 0), 1, 0)
\end{align*}
There are eight positions in the body expression, namely
$\{\epsilon, 1, 2, 3,\allowbreak 11, 12, 111, 112\}$.
The subexpression at position $112$ is $y$, and
$p[u]_{11} = \mathbf{ite}(u \le 0, 1, 0)$.
This corresponds to the decomposition:
\[
(\lambda x,y . x + y, u, \lambda x, y, u . \mathbf{ite}(u \le 0, 1, 0))
\]
\end{example}

With this notation in place, we can formally describe the detection algorithm
in Algorithm~\ref{alg:detect-generic}.

\begin{algorithm}[t]
  \caption{Detection for expression programs.
    \label{alg:detect-generic}}
  \begin{algorithmic}
    \Require association (\assocmeasure{}), influence(\infmeasure{}) measures
    \Procedure{ProxyDetect}{$p, \features, Z, \epsilon, \delta$}
    \State $P \gets \varnothing$
    \For{each term $e$ appearing in $p$}
      \State $p_1 \gets \lambda x_1, \ldots, x_n . e$
      \State $Q \gets \{ q\ |\ p|_q = e \}$
      \For{each $k \in [1, \ldots, |Q|], (q_1, \ldots, q_k) \in Q$}
        \State $p_2 \gets \lambda x_1, \ldots, x_n, u . p[u]_{q_1, \ldots, q_k}$
        \If{$\iota(p_1, p_2) \geq \delta \land d(\llbracket p_1 \rrbracket(\features), Z) \geq \epsilon$}
          \State $P \gets P \cup \{(p_1, p_2)\}$
        \EndIf
      \EndFor
    \EndFor
    \State \Return $P$
    \EndProcedure
\end{algorithmic}
\end{algorithm}

\subsection{Translation}
\label{app:translation}

This section describes the translation of machine learning models
used in our implementation to the term language.

\subsubsection{Decision trees and Rule lists}

\begin{figure}

\centering

\begin{tabular}{c@{\hskip 0pt}c}

\parbox[c]{0.5\columnwidth}{%

\newdimen\nodeDist
\nodeDist=15mm

\begin{tikzpicture}[
    node/.style={%
      draw,      
      font=\small,
      inner sep=2pt,
      anchor=base west
    },
  ]

    \node [node,circle,text width=0.75em] (A) {$x_1$};
    \path (A) ++(-135:\nodeDist) node [node,draw=none] (B) {0};
    \path (A) ++(-45:\nodeDist) node [node,circle,text width=0.75em] (C) {$x_2$};
    \path (C) ++(-135:\nodeDist) node [node,circle,text width=0.75em] (D) {$x_3$};
    \path (D) ++(-135:\nodeDist) node [node,draw=none] (E) {0};
    \path (D) ++(-45:\nodeDist) node [node,draw=none] (F) {1};
    \path (C) ++(-45:\nodeDist) node [node,draw=none] (G) {0};

    \draw (A) -- (B) node [left,pos=0.35,xshift=-1ex] {\footnotesize $\le \frac{1}{2}$}(A);
    \draw (A) -- (C) node [right,pos=0.35,xshift=1ex] {\footnotesize $> \frac{1}{2}$}(A);
    \draw (C) -- (D) node [left,pos=0.35,xshift=-1ex] {\footnotesize $\le 1$}(A);
    \draw (D) -- (E) node [left,pos=0.35,xshift=-1ex] {\footnotesize $\le 0$}(A);
    \draw (D) -- (F) node [right,pos=0.35,xshift=1ex] {\footnotesize $> 0$}(A);
    \draw (C) -- (G) node [right,pos=0.35,xshift=1ex] {\footnotesize $> 1$}(A);
\end{tikzpicture} %
}
&
\parbox[c]{0.4\columnwidth}{%
$
\begin{array}{l@{\hskip 0pt}l@{\hskip 0pt}l}
\mathbf{ite}&(x_1 &\le \frac{1}{2}, \\
&0& \\
&\mathbf{ite}&(x_2 \le 1, \\
&&\mathbf{ite}(x_3 \le 0, 0, 1), \\
&&0)))
\end{array}
$
}
\end{tabular}

\caption{Decision tree and corresponding expression program.}
\label{fig:tree-example}
\end{figure}
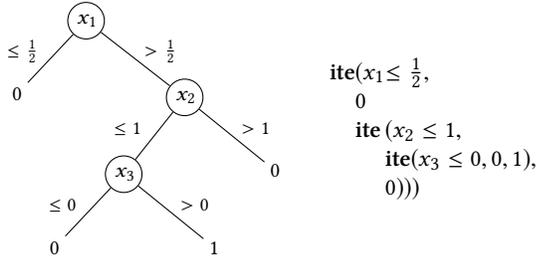

Decision trees can be written in this language as nested
$\mathbf{ite}$ terms, as shown in Figure~\ref{fig:tree-example}.
The Boolean expression in each term corresponds to a guard, and the
arithmetic expressions to either a proper subtree or a leaf. Bayesian
rule lists are a special kinds of decision trees, where the left subtree
is always a leaf.

\subsubsection{Linear models}

Linear regression models are expressed by direct translation into an
arithmetic term, and linear classification models (e.g., logistic
regression, linear support vector machines, Naive Bayes) are expressed
as a single $\mathbf{ite}$ term, i.e.,
\[
  \mathrm{sgn}(\vec{\mathbf{w}}\cdot\vec{\mathbf{x}} + b) \mathrm{\
    becomes\ } \lambda\vec{\mathbf{x}} .
  \mathbf{ite}(\vec{\mathbf{w}}\cdot\vec{\mathbf{x}} + b \ge 0, 1, 0)
\]

Importantly, the language supports $n$-ary operations when
they are associative, and allows for rearranging operands according to
associative and distributive equivalences.
In other words, the language computes on terms modulo an equational
theory. Without allowing such rearrangement, when a linear model is expressed using binary operators, such as
$((((w_1 \times x_1) + (w_2 \times x_2)) + (w_3 \times x_3))$, then
the algorithm cannot select the decomposition:
\[
\begin{array}{l}
  p_1 = \lambda\vec{\mathbf{x}} .
  (w_1 \times x_1) + (w_3 \times x_3) \\
  p_2 = \lambda\vec{\mathbf{x}}, u .
  u + (w_2 \times x_2)
\end{array}
\]

\subsubsection{Decision Forests}

Decision forests are linear models where each linear term is a
decision tree.
We combine the two translations described above to obtain the term
language representation for decision forests.

\subsection{Validity Testing}
\label{app:validity}

We use mutual information to determine the strength of the statistical
association between $\llbracket p_1\rrbracket(\features)$ and $Z$.
Each test of this metric against the threshold $\epsilon$ amounts to a
hypothesis test against a null hypothesis which assumes that
$d(\llbracket p_1\rrbracket(\features), Z) < \epsilon$.
Because we potentially take this measure for each valid decomposition
of $p$, it amounts to many simultaneous hypothesis tests from the same
data source.
To manage the likelihood of encountering false positives, we employ
commonly-used statistical techniques.
The first approach that we use is cross-validation.
We partition the primary dataset $n$ times into training and
validation sets, run Algorithm~\ref{alg:detect-generic} on each
training set, and confirm the reported \proxyuses on the corresponding
validation set.
We only accept reported uses that appear at least $t$ times as valid.

The second approach uses bootstrap testing to compute a $p$-value for
each estimate $\hat{d}(p_1(\features), Z)$, and applying Bonferroni
correction~\cite{dunn1959} to account for the number of simultaneous
hypothesis tests.
Specifically, the bootstrap test that we apply takes $n$ samples of
$(\features, Z)$, $[(\hat{X}_i, \hat{Z}_i)]_{1 \le i \le n}$, and
permutes each $\hat{X}_i, \hat{Z}_i$ to account for the null
hypothesis that $\features$ and $Z$ are independent.
We then estimate the $p$-value by computing:
\[
  p = \frac{1}{n}\sum_{1 \le i \le n} \mathbbm{1}(d(\hat{X}_i,
  \hat{Z}_i) < d(\llbracket p_1 \rrbracket(\features), Z))
\]
After correction, we can bound the false positive discovery rate by
only accepting instances that yield $p \le \alpha$, for sufficiently
small $\alpha$.
We note, however, that this approach is only correct when the
association strength $\epsilon = 1$, as the null hypothesis in this
test assumes that $\llbracket p_1 \rrbracket$ is independent of $Z$.
To use this approach in general, we would need to sample
$[(\hat{X}_i, \hat{Z}_i)]_{1 \le i \le n}$ under the assumption that
$d(\hat{X}_i, \hat{Z}_i) \ge \epsilon$.
We leave this detail to future work.

\section{Algorithms for Repair}
\label{app:repair}

We now provide a formal description of the repair algorithms
informally described in the paper.
Algorithm~\ref{alg:repair}, and~\ref{alg:repair-local} correspond to
~\ref{alg:repair-informal}, and~\ref{alg:repair-local-informal}
respectively.

\begin{algorithm}[t]
\caption{Witness-driven repair.}
\begin{algorithmic}
  \Require association (\assocmeasure{}), influence (\infmeasure{}),
  utility (\utilmeasure{}) measures, oracle ($\oracle{}$)
  \Procedure{Repair}{$p, \features, Z, \epsilon, \delta$} \State
  $ P \gets \set{d \in \Call{ProxyDetect}{p,\features,Z,\epsilon,
      \delta} : \text{ not } \oracle(d)} $ \If{ $ P \neq \emptyset $ }
  \State $(p_1,p_2) \gets \text{element of } P $
  \State
  $p' \gets \Call{ProxyRepair}{p, (p_1,p_2), \features,Z,\epsilon,
    \delta} $ \State \Return
  $\Call{Repair}{p',\features, Z, \epsilon, \delta} $ \Else \State
  \Return $p$
\EndIf
\EndProcedure
\end{algorithmic}
\label{alg:repair}
\end{algorithm}

\begin{algorithm}[t]
\caption{Local Repair.}
\begin{algorithmic}[1]
  \Require association (\assocmeasure{}), influence (\infmeasure{}),
  utility (\utilmeasure{}) measures
  \Procedure{ProxyRepair}{$p, (p_1,p_2), \features, Z, \epsilon,
    \delta$} \State $R \gets \set{}$ \For{each decomp.
    $(p_1',p_2')$ w/ $p_1'$ local to $p_1$ in $p_2$} \State
  $r^* \gets \argmax_{r} \utilmeasure{}\paren{\subst{u}{r}{p_2'}}$
  \State $(p_1'',p_2'') \gets (p_1,p_2) $ with $ r^* $ substituted for
  $ p_1' $
  \If{$\iota(p_1'', p_2'') \leq \delta \lor d(\llbracket p_1''
    \rrbracket(\features), Z) \leq \epsilon$} \State
  $ p^* \gets \subst{u}{r^*}{p_2'} $ \State
  $ R \gets R \cup \set{p^*}$
\EndIf
\EndFor
\State \Return $\argmax_{p^*\in R} \utilmeasure\paren{p^*}$
\EndProcedure
\end{algorithmic}
\label{alg:repair-local}
\end{algorithm}

\subsection{Optimal constant selection}
\label{app:optimal-constant}
As constant terms cannot be examples of \edproxyusage, there is
freedom in their selections as replacements for implicated
sub-programs.
In Algorithm~\ref{alg:repair-local} we pick the replacement that
optimizes some measure of utility of the patched program.
If the given program was constructed as a classifier, we define
utility as the patched program's prediction accuracy on the data set
using 0-1 loss.
Similarly, if the program were a regression model, $v$ would
correspond to mean-squared error.

If the program computes a continuous convex function, as in the case
of most commonly-used regression models, then off-the-shelf convex
optimization procedures can be used in this step.
However, because we do not place restrictions on the functions
computed by programs submitted for repair, the objective function
might not satisfy the conditions necessary for efficient optimization.
In these cases, it might be necessary to develop a specialized
procedure for the model class.
Below we describe such a procedure for the case of decision trees.

\paragraph{Decision trees} Decision trees are typically used for
classification of instances into a small number of classes $C$.
For these models, the only replacement constants that will provide
reasonable accuracy are those that belong to $C$, so in the worst
case, the selection procedure must only consider a small finite set of
candidates.
However, it is possible to calculate the optimal constant with a
single pass through the dataset.

Given a decomposition $(p_1, p_2)$ of $p$, let $\phi$ be the weakest
formula over $p$'s variables such that $\forall \vec{x} .
p_1(\vec{x}) = p(\vec{x})$.
$\phi$ corresponds to the conjoined conditions on the path in $p$
prefixing $p_1$.
We can then define the objective function:
\[
  v(r) = \sum_{\vec{x} \in \vec{X}} \mathbbm{1}(\phi(\vec{x})
  \rightarrow \vec{x}_c = r)
\]
This objective is minimized when $r$ matches the greatest number of
class labels for samples that pass through $p_1$.
This minimizes classification error over $\vec{X}$, and is easily
computed by taking the class-label mode of training samples that
satisfy $\phi$.

\begin{example}
  Consider the tree in Figure~\ref{fig:tree-example}, and assume that
  $x_1$ and $x_2$ are distributed according to
  $\mathcal{N}(\frac{1}{2}, 1)$, and $x_3 = x_1 + x_2$.
  For simplicity, assume that the class label for each instance is
  given exactly by the tree.
  Then given the decomposition:
\begin{align*}
  p_1 &= \lambda \vec{x}.
        \mathbf{ite}(x_3 \le 0, 0, 1) \\
  p_2 &= \lambda \vec{x}, u.
        \mathbf{ite}(x_1 \le 1/2, 0, \mathbf{ite}(x_2 \le 1, u, 0))
\end{align*}
we need to find an optimal constant to replace the subtree rooted at
$x_3$.
In this case, $\phi \defeq x_1 > \frac{1}{2} \land x_2 \le 1$, so we
select
$\vec{X}_\phi = \{\vec{x} \in \vec{X} | x_1 > \frac{1}{2} \land x_2
\le 1\}$ and take the mode of the empirical sample
$[p(\vec{x})]_{\vec{x} \in \vec{X}_\phi}$.
\end{example}

\ifdefined\shortappendix\else

\section{Other experiments}

\begin{figure*}[t]
  \centering
  \includegraphics[width=\linewidth]{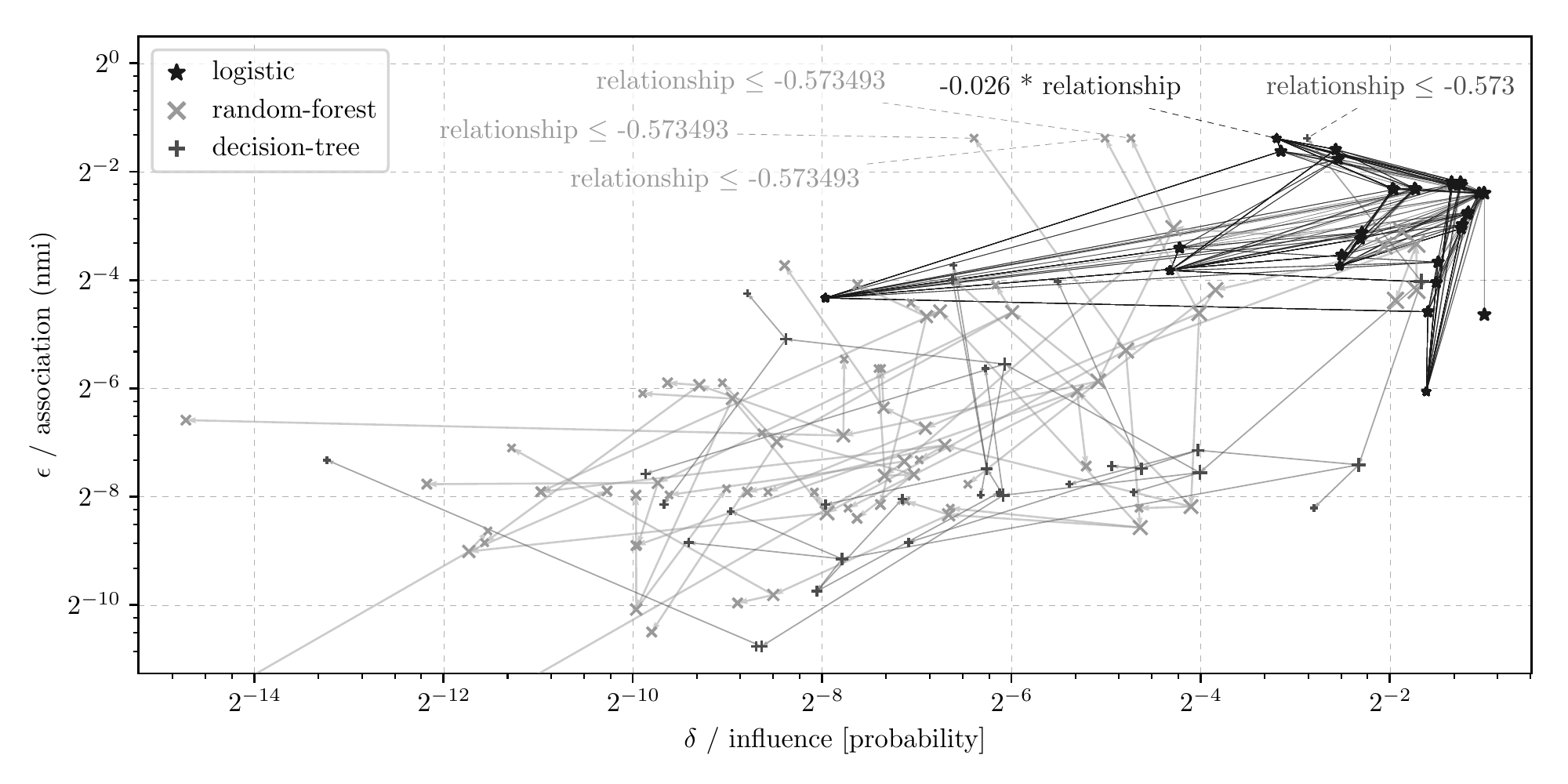}
  \caption{\label{fig:big} The association with
    $\mathtt{marital\_status}$ and influence of the sub-expressions of
    three similarly-sized models trained on the UCI Adult dataset
    (normalized): random forest ($\times$), decision tree ($+$), and
    logistic regression ($\star$).
    The arrows denote the sub-expression relationship among the
    expressions.
    Pointed out are several significant expressions that use the
    $\mathtt{relationship}$ feature.
    Note that in the random forest, the same associated expression
    appears in all three of the trees in that model.
  }
\end{figure*}

\subsection{Details of Case Studies}
\label{app:case-study-details}

\paragraph{Targeted contraception advertising}
We consider a scenario in which a data processor wishes to show
targeted advertisements for contraceptives to females.
To support this goal, the processor collects a dataset from a
randomly-sampled female population containing age, level of education,
number of children, current employment status, income, level of media
exposure, information about the partner's education and occupation,
and the type of contraception used by the individual (if any).
This dataset is used to train a model that predicts whether an
individual uses no contraception, short-term contraception, or
long-term contraception.
This model is then used to determine who to display advertisements to,
under the assumption that individuals who already use short-term
contraception are more likely to be receptive to the advertisements.

Because certain religions ban the use of contraception, users
belonging to such a religion are on the whole less likely to purchase
contraceptives after seeing such an advertisement.
The ad-targeting model does not explicitly use a feature corresponding
to religion, as this information is not available to the system when
ads are displayed.
Furthermore, some users may view the use of this information for
advertising purposes as a violation of their privacy, so the data
processor would like to ensure that the targeting model has not
inferred a proxy for this information that is influential in
determining whether to show an advertisement.

We evaluated this scenario using data collected for the 1987 National
Indonesia Contraceptive Survey~\cite{indonesia-contra}, which contains
the features mentioned above, as well as a feature indicating whether
the individual's religious beliefs were Islam.
To simulate the data processor, we trained a decision tree classifier
to predict contraceptive use over all available features except the
one corresponding to religion.
We then used our detection algorithm to look for a proxy use of
religion, using the available data as ground truth to evaluate the
effectiveness of our approach.

Although this data is representative of a single country, it
illustrates an interesting case of potential use privacy.
Our detection algorithm identified the following intermediate
computation, which was one of the most influential in the entire
model, and the one most closely associated with the religion variable:
$ ite(\mathtt{educ} < 4 \land \mathtt{nchild} \leq 3 \land
\mathtt{age} < 31, \mathtt{no}, \mathtt{yes}) $.
This term predicts that women younger than 31, with below-average
education background and fewer than four children will not use
contraception.
Given that the dataset is comprised entirely of females, closer
examination in fact reveals that just the ``guard'' term
$\mathtt{educ} < 4$ alone is even more closely associated with
religion, and its influence on the model's output is nearly as high.
This reveals that the model is using the variable for education
background as a proxy for religion, which may be concerning given that
this application is focused on advertising.

\paragraph{Student assistance} A current trend in education is the use
of predictive analytics to identify students who are likely to benefit
from certain types of interventions to ensure on-time graduation and
other benchmark goals~\cite{uc-analytics, k12-analytics}.
We look at a scenario where a data processor builds a model to predict
whether a secondary school student's grades are likely to suffer in
the near future, based on a range of demographic features (such as
age, gender, and family characteristics), social information (such as
involvement in extracurricular activities, amount of reported free
time after school), and academic information (e.g., number of reported
absences, use of paid tutoring services, intention to continue on to
higher education).
Based on the outcome of this prediction, the student's academic
advisor can decide whether to pursue additional interventions.

Because of the wide-ranging nature of the model's input features, and
sensitivity towards the privacy rights of minors, the data processor
would like to ensure that the model does not base its decision on
inferred facts about certain types of activities that the student
might be involved with.
For example, alcohol consumption may be correlated with several of the
features used by the model, and it may not be seen as appropriate to
impose an intervention on a student because their profile suggests
that they may have engaged in this activity.
Depending on the context in which such an inference were made, the
processor would view this as a privacy violation, and attempt to
remove it from the model.

To evaluate this scenario, we trained a model on the UCI Student
Alcohol Consumption dataset~\cite{student-alcohol}.
This data contains approximately 700 records collected from Portuguese
public school students, and includes features corresponding to the
variables mentioned above.
Our algorithm found the following proxy for alcohol use:
$ite(\mathtt{studytime} < 2 \land \mathtt{dad\_educ} < 4,
\mathtt{fail}, ...)$, which predicts that a student who spends at most
five hours per week studying, and whose father's level of education is
below average, is likely to fail a class in at least one term.
Further investigation reveals that $\mathtt{studytime} < 2$ was more
influential on the model's output, and nearly as associated with
alcohol consumption, as the larger term.
This finding suggests that this instance of proxy use can be deemed an
appropriate use, and not a privacy violation, as the amount of time a
student spends studying is clearly relevant to their academic
performance.
If instead $\mathtt{dad\_educ} < 4$ alone had turned out to be a proxy
use of alcohol consumption, then it may have been a concerning
inference about the student's behavior from information about their
family history.
Our algorithm correctly identified that this is not the case.

\paragraph{Credit advertisements} We consider a situation where a
credit card company wishes to send targeted advertisements for credit
cards based on demographic information.
In this context, the use of health status for targeted advertising is
a legitimate privacy concern~\cite{dixon14scores}.

To evaluate this scenario, we trained a model to predict interest in
credit cards using the PSID dataset, which contains
detailed demographic information for roughly 10,000 families and
includes features such as age, employment status income, education,
and the number of children.
From this, we trained two models: one that identifies individuals with
student loans and another that identifies individuals with existing
credit cards as the two groups to be targeted.

The first model had a number of instances of proxy use.
One particular subcomputation that was concerning was a subtree of the
original decision tree that branched on the number of children in the
family.
This instance provided negative outcomes to individuals with more
children, and may be deemed inappropriate for use in this context.
In the second model, one proxy was a condition involving income
$\mathtt{income} \leq 33315$.
The use of income in this context is justifiable, and therefore this
may not be regarded as a use privacy violation.

\fi

\ifdefined\shortappendix\else

\subsection{Algorithm Runtime vs. Model Size}
\label{sec:bench-detect-model-size}
\begin{figure*}[t]
  \centering\includegraphics[width=\linewidth]{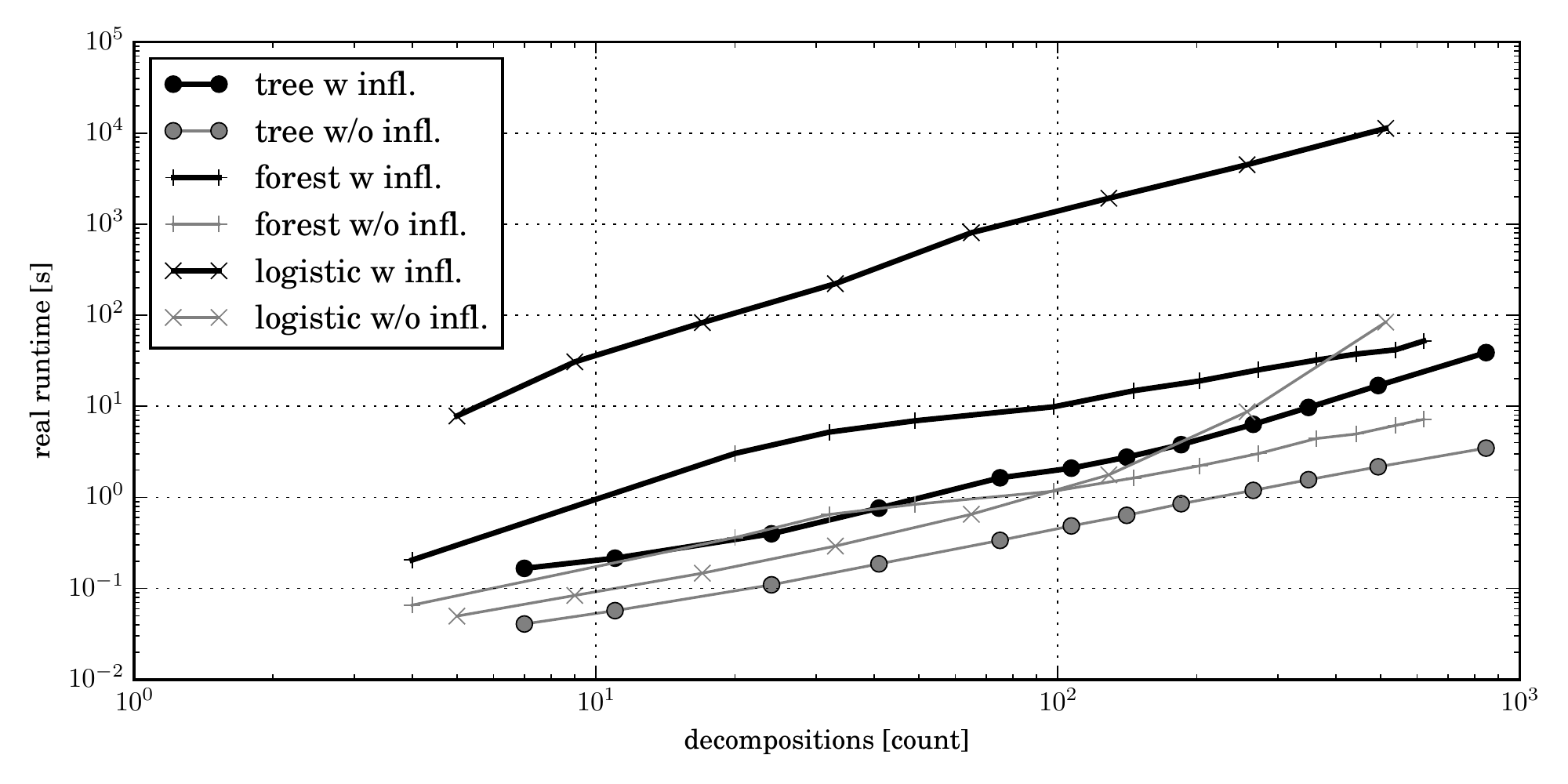}
  \caption{\label{fig:bench-detect-model-size} Detection algorithm
    run-time as a function of the number of decompositions in a
    decision tree($\circ$), a random forest($+$), and a logistic
    regression($\times$) trained on the UCI Adult dataset.
    Detecting violations with $ \epsilon=0.0 $ (black markers and
    line) requires influence computations whereas detecting with
    $ \epsilon=1.0 $ (gray markers and line) does not.
    A 2000 instance sample of the dataset was used with varying
    training depths.}
\end{figure*}

Figure~\ref{fig:bench-detect-model-size} demonstrates the runtime of
the detection algorithm as a function of the number of decompositions
of the analyzed model (a proxy for its size).
We show two trends in that figure.
The black line demonstrates the worst case detection that requires
both association and influence computation for each decomposition
while the gray line demonstrates the best case where only the
association computation is performed.
The runtime in practice would thus fall somewhere between these two
cases, both linear in the number of decompositions.

\fi

\section{Complexity}\label{appendix:complexity}

The complexity of the presented algorithms depend on several factors,
including the type of model being analyzed, the number of elements in
the ranges of sub-programs, and reachability of sub-programs by
dataset instances .
In this section we describe the the complexity characteristics of the
detection and repair algorithms under various assumptions.
Complexity is largely a property of the association and influence
computations and the number of decompositions of the analyzed program.
We begin by noting our handling of probability distributions as
specified by datasets, several quantities of interest, discuss the
complexity of components of our algorithms, and conclude with overall
complexity bounds.

\subsection{Distributions, datasets, and
  probability}\label{sec:complexity-distributions}

It is rarely the case that one has access to the precise distribution
from which data is drawn.
Instead, a finite sample must be used as a surrogate when reasoning
about random variables.
In our formalism we wrote $ X \samp \population $ to designate
sampling of a value from a population.
Given a dataset surrogate $ \dataset $, this operation is implemented
as an enumeration $ \mathbf{x} \in \dataset $, with each element
having probability $ 1 / \setsize{\dataset} $.
We will overload the notation and use $ \dataset $ also as the random
variable distributed in the manner just described.
We assume here that the sensitive attribute $ Z $ is a part of the
random variable $ X $.

The following sections use the following quantities to express
complexity bounds, mostly overloading prior notations:
\begin{itemize}
\item{} $ \datasize $ - The number of instances in the population
  dataset.
\item{} $ \progsize $ - The number of expressions
  in a program $ \prog $ being analyzed.
\item{} $ \zs $ - The number of elements in the support of
  $ Z $.
\item{} $ k $ - The maximum number of unique elements in support of
  every sub-expression, that is
  $ \max_{p' \in \prog} \setsize{\support{\denote{p'} \dataset}} $.
\item{} $ c $ - The number of decompositions in a given program.
  We will elaborate on this quantity under several circumstances later
  in this section.
\item{} $ b $ - The minimum branching factor of sub-expressions in a
  given program.
\end{itemize}

We will assume that the number of syntactic copies of any
sub-expression in a program is no more than some constant.
This means we will ignore the asymptotic effect of decompositions with
multiple copies of the same sub-program $ p_1 $.

The elementary operation in our algorithms is a lookup of a
probability of a value according to some random variable.
We pre-compute several probabilities related to reachability and
contingency tables to aid in this operation.
When we write ``$p_1$ is reached'', we mean that the evaluation of
$ p $, containing $ p_1 $, on a given instance $ \features $, will
reach the sub-expression $p_1$ (or that $p_1$ needs to be evaluated to
evaluate $p$ on $\features$).

\paragraph*{Probability pre-computation}\hfill

\noindent For every decomposition
$ \denote{p_2}\paren{X, \denote{p_1}X} = \denote{p}\paren{X} $, we
compute:

\begin{enumerate}
\item{} the r.v.
  $ (\denote{p_1} \features, \features_Z) $ for
  $ \features \samp \dataset $,
\item{} the r.v.
  $ \features | \paren {p_1 \text{ is reached by } \features} $ for
    $ \features \samp \dataset $, and
\item{} the value
  $ Pr_{\features \samp \dataset}\paren{ p_1 \text{ is reached by } \features}$.
\end{enumerate}

\vspace{0.25cm}

In point (1) above we write $ \features_Z $ to designate the sensitive
attribute component of $ \features $, hence this point computes the
r.v.
representing the output of $ p_1 $ along with the sensitive attribute
$ Z $.
This will be used for the association computation.

The complexity of these probability computations varies depending on
circumstances.
In the worst case, the complexity is
$ \bigO{c \datasize \progsize} $.
However, under some assumptions related to programs $ \prog $ and
datasets $ \dataset $, these bounds can be improved.
We define two types of special cases which we call splitting and
balanced:

\begin{definition}$ \prog $ is \emph{splitting} for $ \dataset $ iff
  it has at most a constant number of reachable op operands (arguments of
  $\text{op}$ expressions).
\end{definition}

The Decision trees are local for any dataset as they do not contain
any op operands (they do contain relop operands).
Further, if number of trees in random forests or number of
coefficients in linear regression are held constant, then these models
too are splitting for any dataset.
The reasoning behind this definition is to prohibit arbitrarily large
programs that do not split inputs using if-then-else expressions.
It is possible to create such programs using arithmetic and boolean
operations, but not using purely relational operations.

\begin{definition} $ \prog $ is \emph{$b$-balanced} for $ \dataset $ iff
  all but a constant number of sub-expressions $ e'$ have parent $ e $
  with $ b > 1 $ sub-expressions which split the instances that reach
  them approximately equally among their children.
\end{definition}

Balanced implies splitting as op operands do not satisfy the balanced
split property hence there has to be only a constant number of them.
Also, the definition is more general than necessary for the language
presented in this paper where the branching factor is always $ 2 $
because the if-then-else expressions are the only ones that can
satisfy the balanced split condition.
Decision trees trained using sensible algorithms are usually balanced
due to the branch split criteria employed preferring approximately
equal splits of training instances.
For the same reason, if the number of trees are held constant,
then random forests are also likely to be balanced.

When $ p $ is splitting for $ \dataset $, the probability computation
step reduces to $ \bigO{\datasize \progsize^2} $.
This stems from the fact that the number of decompositions is
asymptotically equal to the number of sub-expressions (limits to
operands prevent more decompositions).
Further, if $ p $ is $b$-balanced for $ \dataset $, the probability
pre-computation reduces to $ \bigO{\datasize \log_b \datasize} $.
In the language presented $ b = 2 $.
These bounds derive similarly to the typical divide and conquer
program analysis; there are $ \log_b \datasize $ layers of
computation, each processing $ \datasize $ instances.

\subsection{Influence and Association}\label{sec:estimates}

Our proxy definition further relies on two primary quantities used in
Algorithm~\ref{alg:detect-generic-informal}, influence and
association.
We describe the methods we use to compute them here.

\paragraph{Quantitative decomposition influence}

Given a decomposition $(p_1, u, p_2)$ of $p$, the influence of $p_1$
on $p_2$'s output is defined as:
\begin{equation*}
  \iota(p_1, p_2) \defeq \expectsub{X, X' \samp
    \dataset}{\prob{\denote{p_2}\paren{\features,
      \denote{p_1}{\features}} \ne \denote{p_2}\paren{\features,
      \denote{p_1}{\features'}}}}
\end{equation*}

This quantity requires $ \datasize^2 $ samples to compute in general.
Each sample takes at most $ \bigO{\progsize} $ time, for a total of
$ \bigO{\progsize \datasize^2} $.
However, we can take advantage of the pre-computations described in
the prior section along with balanced reachability criteria and
limited ranges of values in expression outputs to do better.
We break down the definition of influence into two components based on
reachability of $ p_1 $:
\begin{align*}
  \iota(p_1, p_2)
  &\defeq \expectsub{\features, \features' \samp
    \dataset}{\Pr\paren{\denote{p_2}\paren{\features,
    \denote{p_1}{\features}} \neq \denote{p_2}\paren{\features,
    \denote{p_1}{\features'}}}} \\
  &= \expectsub{\features \samp \dataset}
    {\expectsub{\features'\samp \dataset}
    {\Pr\paren{\denote{p_2}\paren{\features,
    \denote{p_1}{\features}} \neq \denote{p_2}\paren{\features,\denote{p_1}{\features'}}}}} \\
  &= \prob{p_1 \text{ not reached}}\cdot \expectsub{\features \samp
    \dataset|p_1 \text{ not reached}}{\cdots} \\
  & \;\;\;\; + \prob{p_1 \text{ reached}} \cdot \expectsub{\features \samp \dataset|p_1 \text{ reached}}{\cdots} \\
  & = 0 + \prob{p_1 \text{ reached}} \cdot \\
  & \;\;\;\; \expectsub{\features \samp \dataset|p_1 \text{
    reached}}{\expectsub{\features \samp \dataset}{\Pr\paren{\denote{p}(\features) \neq
    \denote{p_2}\paren{\features, \denote{p_1}{\features'}}}}} \\
  & = \prob{p_1 \text{ reached}} \cdot \\
  & \;\;\;\; \expectsub{\features \samp \dataset|p_1 \text{ reached}}{\expectsub{\mathbf{Y} \samp \denote{p_1}\dataset}{\Pr\paren{\denote{p}(\features) \neq \denote{p_2}\paren{\features, \mathbf{Y}}}}}
\end{align*}
Note that all both random variables and one probability value in the
final form of influence above have been pre-computed.
Further, if the number of elements in the support of
$ \denote{p_1} X $ is bounded by $ k $, we compute influence using
$ k \datasize $ samples (at most $ \datasize $ for $ \features $ and
at most $ k $ for $ \mathbf{Y}$), for total time of
$ \bigO{k \progsize \datasize} $.

Influence can also be estimated, $ \hat{\iota} $ by taking a
\emph{sample} from $\dataset \times \dataset$.
By Hoeffding's inequality~\cite{Hoeffding:1963}, we select the
subsample size $n$ to be at least $\log(2/\beta)/2\alpha^2$ to ensure
that the probability of the error
$\hat{\iota}(p_1, p_2) - \iota(p_1, p_2)$ being greater than $\beta$
is bounded by $\alpha$.

\paragraph{Association}

As discussed in Section~\ref{sec:defn}, we use mutual information to
measure the association between the output of a subprogram and $Z$.
In our pre-computation steps we have already constructed the r.v.
$ \paren{\denote{\prog_1}\features, \features_Z} $ for
$ \features \samp \dataset $.
This joint r.v.
contains both the subprogram outputs and the sensitive attribute hence
it is sufficient to compute association metrics.
In case of normalized mutual information, this can be done in time
$ \bigO{ k \zs} $, linear in the size of the support of this random
variable.

\subsection{Decompositions}

The number of decompositions of a model determines the number of
proxies that need to be checked in detection and repair algorithms.
We consider two cases, splitting and non-splitting programs.
For splitting models, the number of decompositions is bounded by the
size of the program analyzed, whereas in case of non-splitting models,
the number of decompositions can be exponential in the size of the
model.
These quantities are summarized in Table~\ref{table:decompositions}.

\begin{table}[ht]
\centering
\begin{tabular}{|p{1.75in}|c|c|}\hline
  & splitting            & non-splitting     \\ \hline
  worst-case general                           & $ \bigO{\progsize} $ & $ \bigO{2^{\progsize}}     $ \\ \hline
  linear model with (constant or $ f $ ) number of coefficients         & $ \bigO{1}     $ & $ \bigO{2^f}     $ \\ \hline
  decision tree of height $ h $                & $ \bigO{2^h}   $ & $ \bigO{2^h}     $ \\ \hline
  random forest of (constant or $ t $) number of trees of height $ h $ & $ \bigO{2^h} $ & $ \bigO{2^t 2^h} $ \\ \hline
\end{tabular}
\vspace{0.5cm}
\caption{\label{table:decompositions}
  The number of decompositions in various types of models.
  When we write ``(constant or f)'' we denote two cases: one in which
  a particular quantity is considered constant in a model making it
  satisfy the splitting condition, and one in which that same quantity
  is not held constant, falsifying the splitting condition.
}
\end{table}

\subsection{Detection}
The detection algorithm can be written $\bigO{A + B \cdot C}$, a
combination of three components.
$ A $ is probability pre-computation as described earlier in this
section, $ B $ is the complexity of association and influence
computations, and $ C $ is the number of decompositions.

\begin{table}
\begin{tabular}{|l|l|}\hline
non-splitting & $ \bigO{\progsize c \paren{\datasize + k \datasize}} $ \\ \hline
splitting     & $ \bigO{\progsize \paren{\datasize \progsize + c k \datasize}}$ \\ \hline
$b-$balanced  & $ \bigO{\datasize \log_b \datasize + c k \progsize \datasize}$   \\ \hline
\end{tabular}
\vspace{0.75cm}\caption{\label{table:overall}
  The complexity of the detection algorithm under various conditions, as a function of the number of decompositions.}
\end{table}

The complexity in terms of the number of decompositions under various
conditions is summarized in Table~\ref{table:overall}.
Instantiating the parameters, the overall complexity ranges from
$ \bigO{\datasize \log_b \datasize + \progsize^2 \datasize} $ in case of
models like balanced decision trees with a constant number of classes,
to
$ \bigO{\progsize 2^{\progsize} \datasize^2}
$ in models with many values and associative expressions like linear
regression.
If the model size is held constant, these run-times become
$ \bigO{\datasize \log_b \datasize} $ and $ \bigO{\datasize^2} $,
respectively.

\end{document}